\documentclass[aps,prl,reprint,longbibliography, asmath, amssymb]{revtex4-2}

\usepackage{graphicx}
\usepackage{dcolumn}
\usepackage{bm}
\usepackage[thinc]{esdiff}
\usepackage{mathtools}
\usepackage{natbib}
\usepackage{subfigure}
\usepackage{wrapfig}
\usepackage{xcolor}

\newsavebox{\astrutbox}
\sbox{\astrutbox}{\rule[-5pt]{0pt}{20pt}}

\usepackage{amsmath} 
 \usepackage{amssymb,euscript,graphicx}

\newcommand{\beq}{\begin{equation}}
\newcommand{\eeq}{\end{equation}}
\newcommand{\barr}{\begin{array}}
\newcommand{\earr}{\end{array}}
\newcommand{\beqarr}{\begin{eqnarray}}
\newcommand{\eeqarr}{\end{eqnarray}}
\newcommand{\beqar}{\begin{eqnarray*}}
\newcommand{\eeqar}{\end{eqnarray*}}
\newcommand{\bef}{\begin{figure}}
\newcommand{\eef}{\end{figure}}

\def\XXint#1#2#3{{\setbox0=\hbox{$#1{#2#3}{\int}$}
     \vcenter{\hbox{$#2#3$}}\kern-.5\wd0}}

\begin{document}

\preprint{APS/123-QED}

\title[Variational Projection of Navier-Stokes]{Variational Projection of Navier-Stokes: \\
Fluid Mechanics as a Quadratic Programming Problem}

\author{Haithem Taha}\email{hetaha@uci.edu}
\author{Kshitij Anand}

\affiliation{University of California, Irvine}


\begin{abstract}
Gauss's principle of least constraint transforms a dynamics problem into a pure minimization problem, where the total magnitude of the constraint force is the cost function, minimized at each instant. A candidate motion whose evolution minimizes the Gaussian cost at each instant is guaranteed to satisfy Newton's equation. In other words, Newton's equation is the first-order necessary condition for minimizing the Gaussian cost, subject to the given kinematic constraints. The principle of minimum pressure gradient (PMPG) is to incompressible fluid mechanics what Gauss's principle is to particle mechanics. The PMPG asserts that an incompressible flow evolves from one instant to another by minimizing the $L^2$-norm of the pressure gradient force. A candidate flow field whose evolution minimizes the pressure gradient cost at each instant is guaranteed to satisfy the Navier-Stokes equation. Consequently, the PMPG transforms the incompressible fluid mechanics problem into a pure minimization framework, allowing one to determine the evolution of the flow field by solely focusing on minimizing the cost---without directly invoking the Navier-Stokes equation. In this paper, we show that the resulting minimization problem is a convex Quadratic Programming (QP) problem---one of the most computationally tractable classes in nonlinear optimization, which has a rich literature with many efficient algorithms. Moreover, leveraging tools from analytical mechanics and the Moore-Penrose theory of generalized inverses, we derive an analytical solution for this QP problem. As a result, we present an explicit formula for the projected dynamics of the spatially discretized Navier-Stokes equation on the space of divergence-free fields. The resulting ordinary differential equation (ODE) is ready for direct time integration, eliminating the need for solving the Poisson equation in pressure at each time step. It is typically an explicit nonlinear ODE with constant coefficients. This compact form is expected to be highly valuable for both simulation and theoretical studies, including stability analysis and flow control design. We demonstrate the framework on the benchmark problem of unsteady flow in a lid-driven cavity.
\end{abstract}

\maketitle

\section{Introduction}
Fluid mechanics, like solid mechanics, is a fundamental branch of mechanics. The parent discipline, mechanics, has two primary approaches: (i) Newtonian mechanics and (ii) analytical or variational mechanics. The former approach is encapsulated in the famous equation $F=ma$. In contrast, analytical mechanics is less likely to be epitomized in a simple formula; it subsumes different approaches: (a) Lagrangian mechanics, (b) Hamiltonian mechanics, (c) variational mechanics, among other variants \citep{Lanczos_Variational_Mechanics_Book}. The central concept in analytical mechanics is that the dynamics of the entire system, encompassing all degrees of freedom, can be derived from a single scalar function. This scalar function encodes all the dynamical information about the system. This scalar is the Lagrangian function in Lagrangian mechanics, the Hamiltonian function in Hamiltonian mechanics, and the cost function (or functional) in variational mechanics.

The framework of analytical mechanics, in all its variants, has proven  significantly more effective than the Newtonian framework in tackling challenging problems in mechanics; in both general relativity and quantum mechanics, the Newtonian formulation failed, while analytical mechanics was flexible enough to adapt to these new paradigms \citep{Lanczos_Variational_Mechanics_Book}. Moreover, even within the realm of classical mechanics scales---where the Newtonian framework is sufficient---the analytical mechanics framework is notably more efficient in handling constrained mechanical systems \citep{Papastavridis}, particularly those with \textit{holonomic} constraints. For example, consider a mechanical system with $N$ particles in three-dimensions, resulting in $3N$ degrees of freedom. However, they are not all independent; consider, in addition, $c$ holonomic constraints; i.e., $c$ relations that form a $(3N-c)$-dimensional manifold on which the dynamics takes place. In Newtonian mechanics, $3N+c$ equations must generally  be solved to determine the motion: the $3N$ degrees of freedom in addition to the $c$ constraint forces required to maintain the assigned constraints. In contrast, only $3N-c$ equations are required within the framework of analytical mechanics to describe the motion on the configuration manifold; constraint forces do not appear in this case, as the dynamics is projected onto the configuration manifold, where all holonomic constraints are inherently satisfied, and constraint forces are normal to the configuration manifold. Hence, for a large number of constraints, the analytical mechanics formulation is significantly more efficient.

Based on the above discussion, we anticipate that the analytical mechanics framework will prove effective in addressing some of the challenges within the sub-branch of fluid mechanics, particularly for incompressible flows where the dynamics is constrained by the divergence-free condition on the velocity field---equivalently, the volume-preserving constraint on the flow map. A flow map $\Phi_t:\mathbb{R}^3\to\mathbb{R}^3$ is a map that takes an initial flow configuration (Lagrangian coordinates) to its final configuration (spatial coordinates) after evolving under the flow dynamics for time $t$. This map is a diffeomorphism on $\mathbb{R}^3$ (i.e., a smooth invertible map from $\mathbb{R}^3$ to $\mathbb{R}^3$). In particular, for incompressible flows, it is a volume-preserving diffeomorphism. Hence, the incompressibility constraint restricts the dynamics to the manifold of volume-preserving diffeomorphisms, consisting of all diffeomorphisms on $\mathbb{R}^3$ with a unit Jacobian determinant. A tangent vector to this manifold corresponds to a divergence-free velocity field in the actual space. By the Helmholtz–Hodge decomposition, the pressure gradient force $\nabla p$ (or any curl-free vector) is orthogonal to this manifold \citep{Chorin_Marsden_Book,Geometric_Control_Fluid_Dynamics}. For incompressible flows, the pressure force acts as the constraint force that ensures the continuity constraint \citep{Pressure_BCs,Chorin_Marsden_Book,Pressure_Constraint_General,Morrison2020lagrangian}.

Given this background, it is anticipated that a proper analytical mechanics formulation for incompressible flows will handle the system directly on the configuration manifold (of volume preserving diffeomorphisms), thereby eliminating the constraint force (i.e., the pressure force) from the picture, since it is orthogonal to the manifold. If successful, such a formulation is expected to be more efficient than the Newtonian framework, which explicitly includes the pressure force in the governing equation---the Navier-Stokes equation. This fact is well-known to fluid mechanicians, though its practical implementation remains unclear in general scenarios. For instance, it is widely recognized that projecting the Navier-Stokes equation onto the space of divergence-free velocity fields eliminates the pressure term \citep[e.g.,][]{Chorin_Projection,Temam_Projection,Moin_Incompressible1,Pressure_BCs,Chorin_Marsden_Book}. However, the standard technique for achieving such a projection requires solving the Poisson equation in pressure at every time step \citep{Chorin_Projection,Temam_Projection,Moin_Incompressible1}, which is the most computationally expensive step in simulating turbulent flows \citep{Poisson_Equation_Pressure_Cost}.

One of the early pioneering theoretical efforts in the development of an analytical mechanics formulation of incompressible flows was by Vladimir Arnold who showed that the motion of ideal fluids follows geodesics (i.e., straight lines) on the manifold of volume preserving diffoemorphisms \citep{Arnold_French}. This result is perfectly aligned with concepts in analytical mechanics, established by Jacobi and Darboux (see Dugas \citep{Dugas}): since ideal flows are forceless except for the pressure force, which is a constraint force orthogonal to their configuration manifold, it implies that they are \textit{free} on the configuration manifold. A free mechanical system evolves along geodesics of its configuration manifold \citep{Lanczos_Variational_Mechanics_Book,Bullo_Lewis_Book,Geometric_Control_Fluid_Dynamics}---a consequence of the principle of least action. However, because the principle of least action does not necessarily apply to dissipative systems, extending Arnold's result to real fluids with dissipative viscous forces has not yet been possible.

There have been various efforts to develop variational formulations for the Navier-Stokes equations by extending the principle of least action to account for dissipative forces \citep{Variational_Principles_Fluids_Stochastic,Variational_Principles_NS,Variational_Principles_Fluids_Stochastic_Gomes,Variational_Principles_Fluids_Stochastic2,Variational_Principles_NS_Nonholonomic,Variational_Principles_NS_2n,Variational_Principles_NS_Nonholonomic2,DeVoria_Hamiltonian_JFM}. The primary goal of these efforts is to construct a variational formulation whose first-order necessary condition yields the Navier-Stokes equation. However, because all of these efforts rely on variants of the principle of least action---which does not inherently accommodate dissipative forces---the resulting formulations may appear unnatural. Instead of confining the dynamics to the configuration manifold of volume-preserving diffeomorphism, they consider the dynamics embedded within an even larger space. For instance, Galley's framework \cite{Nonconservative_Forces} converts a dissipative $n$-degree-of-freedom system into a conservative $2n$-degree-of-freedom one, where the energy dissipated from the original $n$-copy is fed into the fictitious $n$-copy, resulting in a larger, but conservative system that is amenable to the principle of least action. In a similar spirit, the framework of Sanders et al. \citep{DeVoria_Hamiltonian_JFM} transforms a generic $n^{\rm{th}}$-order differential equation (which may not originate from a variational formulation) into a $2n^{\rm{th}}$-order equation that possesses a Hamiltonian structure. Though mathematically elegant, the practical value of these variational formulations remains to be tested.

Gauss's principle of least constraint \citep{Gauss_Least_Constraint} stands in a stark contrast to the principle of least action due to its natural ability to directly account for dissipative forces. Unlike the principle of least action, Gauss's principle imposes no restrictions on the nature of impressed forces \citep{Papastavridis,Udwadia_Kalaba_Book}. Consequently, it can handle dissipative systems directly on their natural configuration manifold without embedding in higher dimensions or introducing stochasticity. Another major distinction between Gauss's principle of least constraint and Hamilton's principle of least action \footnote{The principle of least action is indeed due to Maupertuis \citep{Dugas}, but the currently prevalent form of the principle is due to \cite{Hamilton}.} is that the former is a true minimum principle, whereas the latter is merely a stationary principle---the term ``least" is a misnomer. Therefore, Gauss's principle essentially converts the dynamics problem into a pure minimization problem where standard optimization techniques can be employed. In particular, if the constraints (holonomic or nonholonomic) are linear in accelerations \footnote{Constraints that are nonlinear in position or velocity are, in fact, linear in accelerations.}, which encompasses the overwhelming majority of scenarios in mechanics, the resulting optimization problem is a convex quadratic programming (QP) problem, which is arguably the most tractable class in nonlinear optimization \citep{QP_Nonlinear_Programming_Book,QP_Boyd_Book}. Moreover, Udwadia and Kalaba \citep{Udwadia_Kalaba_Original}, managed to obtain an analytical solution to this minimization problem induced by Gauss's principle in finite dimensions, resulting in ODEs that can be directly marched forward in time without requiring iterative computations for Lagrange multipliers to enforce the constraints. This approach has recently become popular in the multi-body dynamics and robotics literature \citep{UK_MB,UK_MB2,UK_MB3,UK_Robotics,UK_Robotics2,Udwadia_Trajectory_Tracking_Application,Udwadia_Kalaba_Review}

Gauss's principle was recently extended to incompressible fluid mechanics, leading to the Principle of Minimum Pressure Gradient (PMPG) \citep{PMPG_PoF}. This principle asserts that an incompressible flow evolves from one time instant to another by minimizing the total magnitude of the pressure gradient force over the domain. It has been proved that the Navier-Stokes equation is the first-order necessary condition for minimizing the pressure gradient cost subject to the continuity constraint. In essence, the PMPG is to incompressible fluid mechanics what Gauss's principle is to particle mechanics. As such, the PMPG converts a fluid mechanics problem into a pure minimization one, in contrast to the many existing variational principles of fluid mechanics that rely on the principle of stationary action \citep{Variational_Principles_Fluids_Stochastic,Variational_Principles_NS,Variational_Principles_Fluids_Stochastic_Gomes,Variational_Principles_Fluids_Stochastic2,Variational_Principles_NS_Nonholonomic,Variational_Principles_NS_2n,Variational_Principles_NS_Nonholonomic2}. Moreover, since the continuity constraint is linear, the resulting optimization problem in terms of the local acceleration has a convex quadratic cost subject to linear constraint. Consequently, any standard discretization scheme yields a convex quadratic programming (QP) problem, analogous to that encountered when applying Gauss's principle to particle mechanics whose analytical solution was obtained by Udwadia and Kalaba \citep{Udwadia_Kalaba_Original,Udwadia_Kalaba_Review}.

In this paper, we aim to achieve two goals. First, we will demonstrate how the PMPG transforms an incompressible fluid mechanics problem into a convex QP problem, which is widely recognized as one of the most tractable classes of nonlinear optimization. Such problems have a unique (global) minimum and are supported by an extensive body of literature detailing efficient computational techniques \citep{QP_Nonlinear_Programming_Book,QP_Boyd_Book}, commercial packages (e.g., MOSEK, Gurobi, and IBM CPLEX), and open-source software (e.g., IPOPT). Unlike general nonlinear problems, which can be NP-hard or undecidable, convex QP is solvable in polynomial time, making it computationally practical for large-scale problems. In this framework, the convex QP formulation eliminates the need to solve a Poisson equation for pressure. That is, instead of solving the Poisson equation to determine the acceleration $\frac{\partial \bm{u}}{\partial t}$ at each time step, a convex QP problem will be solved directly to obtain $\frac{\partial \bm{u}}{\partial t}$. Second, we will adapt Udwadia and Kalaba's solution \citep{Udwadia_Kalaba_Original} of the QP problem arising from applying Gauss's principle to particle mechanics to obtain an analytical solution of the QP problem induced by the application of the PMPG to incompressible flows. As such, we will derive an ODE that can be directly marched forward in time without requiring iterations or solving Poisson's equation for pressure. Fundamentally, this ODE represents the projection of the Navier-Stokes equation onto the space of divergence-free fields. Its compact form is anticipated to be of great benefit to fluid mechanicians for stability analysis and flow control, as it is directly amenable to tools from nonlinear systems theory \citep{Khalil}.


\section{Gauss's Principle of Least Constraint}
Consider the dynamics of $N$ particles, each of mass $m_i$ and inertial acceleration $\bm{a}_i$ subject to impressed forces $\bm{F}_i$ and constraint forces $\bm{R}_i$. Newton's equation can be written for each particle as:
\begin{equation}\label{eq:Newton}
  m_i \bm{a}_i = \bm{F}_i + \bm{R}_i \;\;\; \forall i=1,..,N,
\end{equation}
where the decomposition of forces on the right hand side follows the standard classification in analytical mechanics \citep{Lanczos_Variational_Mechanics_Book,Papastavridis}, dividing them into:  (i) impressed forces $\bm{F}_i$, which are known (either directly or through  constitutive laws) applied forces (e.g., gravitational, elastic, viscous); and (ii) constraint forces $\bm{R_i}$ whose raison d'etre is to enforce kinematical or geometrical constraints (e.g., reaction forces in solid mechanics); if a constraint is removed, the corresponding constraint force ceases to exist. These forces do not contribute to the motion that satisfies the constraints; their primary purpose is to maintain the constraints (i.e., prevent any deviation from them). Examples include the force in a pendulum rod, the normal force acting on a particle sliding over a surface, and the reaction force at a hinge.

Gauss postulated his principle in a seminal four-page paper \citep{Gauss_Least_Constraint}, where he defined \textit{free motion} as the motion that would occur in the absence of constraints; i.e., it is simply given by $\bm{a}_i^{\rm{free}} = \frac{\bm{F}_i}{m_i}$. He then proposed the following principle to determine the constrained (i.e., actual) motion:
\begin{center}
``\textit{The motion of a system of $N$ material points takes place in every moment in maximum accordance with the free movement or under least constraint, the measure of constraint, is considered as the sum of products of mass and the square of the deviation to the free motion.}"
\end{center}
That is, Gauss's principle implies that the motion takes place such that the quantity
\begin{equation}\label{eq:Gauss}
  Z = \frac{1}{2} \sum_{i=1}^N m_i\left( \bm{a}_i- \frac{\bm{F}_i}{m_i} \right)^2
\end{equation}
is minimized at every instant, provided that the constraints are satisfied.

The principle is intriguingly intuitive. In the absence of constraints (i.e., no constraint forces), the actual motion coincides with the free motion. However, in the presence of constraints, is it quite intuitive to expect that the actual motion will be the closest one---satisfying the constraints---to the free motion. In other words, among all kinematically admissible motions (i.e., those satisfying the constraints), the actual motion is the one with the least instantaneous deviation from the free motion. Note that the quantity $Z$ can be expressed as $Z = \frac{1}{2} \sum_{i=1}^N \frac{1}{m_i}\bm{R}_i^2$, i.e., the sum of the squares of the magnitudes of the constraint forces must be minimum---hence the name least constraint.

Consider a system of $N$ particles in three dimensions (i.e., $\bm{a}=[\bm{a}_1^T, ..., \bm{a}_N^T]^T\in\mathbb{R}^{3N}$) subject to $c< 3N$ constraints that may be arbitrarily nonlinear in positions and velocities:
\begin{equation}\label{eq:Constraints_General}
\Psi_k(\bm{x},\bm{v}) = 0, \; k=1,...,c,
\end{equation}
where $\bm{x}\in\mathbb{R}^{3N}$ and $\bm{v}\in\mathbb{R}^{3N}$ are arrays of positions and velocities of the $N$ particles, respectively. Differentiating Eq. (\ref{eq:Constraints_General}) with respect to time yields a form that is linear in accelerations:
\begin{equation}\label{eq:Constraints_Linear}
[A_{kj}(\bm{x},\bm{v})] a_j = b_k
\end{equation}
for some $\bm{A}\in\mathbb{R}^{c\times3N}$, $\bm{b}\in\mathbb{R}^c$. To solve this dynamics problem using Newton's approach, the $3N+c$ equations (\ref{eq:Newton},\ref{eq:Constraints_Linear}) must be solved at each time step to determine the $3N$ accelerations $\bm{a}$ and the $c$ constraint forces.

Gauss's principle transforms this dynamics problem into the following minimization problem: minimize the Gaussian cost $Z$ in Eq. (\ref{eq:Gauss}) over $\bm{a}\in\mathbb{R}^{3N}$, subject to the constraint (\ref{eq:Constraints_Linear}):
\begin{equation}\label{eq:Gauss_QP_Problem}
\min_{\bm{a}} \; Z(\bm{a}) \;\;\; \mbox{s.t.} \;\;\; [\bm{A}(\bm{x},\bm{v})] \bm{a} = \bm{b},
\end{equation}
which is a convex QP problem. The solution of this optimization problem has $3N$ variables $\bm{a}$ and $c$ Lagrange multipliers, which are required to enforce the $c$ constraints (\ref{eq:Constraints_Linear}). That is, while Newton's formulation of dynamics provides an explicit equation for the accelerations $\bm{a}$ (in terms of impressed and constraint forces) which, in turn, dictates the evolution of motion at each instant, Gauss's formulation represents dynamical evolution as an instantaneous minimization problem in which the accelerations are the optimization variables. The optimal accelerations---resulting from this instantaneous minimization problem---uniquely determine the motion's evolution at each instant. This formulation is fundamentally different from common variational formulations in analytical mechanics, which treat dynamical evolution as a single optimization problem with a cost function defined as a time integral over the excursion, and the optimization variable being the entire trajectory of the motion. Such a formulation essentially requires calculus of variations, even for particle mechanics. See Section II and Table I in \cite{PMPG_PoF} for a detailed comparison between Gauss's formulation and Lagrangian mechanics using least action.

The first-order necessary condition for this constrained optimization problem is obtained by augmenting the cost $Z$ with Lagrange multipliers $\bm\lambda\in\mathbb{R}^c$ to enforce the constraint (\ref{eq:Constraints_Linear}):
\[ \mathcal{L}(\bm{a},\bm\lambda) = Z(\bm{a}) + \sum_{k=1}^c \lambda_k \left(\sum_{j=1}^{3N} A_{kj} a_j - b_k\right), \]
and then setting the derivative of the Lagrangian $\mathcal{L}$ with respect to each entry $a_j$ in the array $\bm{a}$ of accelerations to zero:
\begin{equation}\label{eq:FONC}
\frac{\partial \mathcal{L}}{\partial a_j}= 0 \;\Rightarrow \; m_j a_j = F_j +\sum_{k=1}^c \lambda_k A_{kj},
\end{equation}
where $m_j$ here is the mass of the particle associated with $a_j$ (i.e., $a_j$ is one of its acceleration components), and $F_j$ is the corresponding component of the impressed force. The first-order necessary condition (\ref{eq:FONC}) is identical to Newton's equation (\ref{eq:Newton}), with the identification that the $j^{\rm{th}}$ entry of the array $\bm{R}$ of constraint forces is written as
\[ R_j = \sum_{k=1}^c \lambda_k A_{kj}. \]
Thus, Newton's equation represents the first-order necessary condition of the optimization problem induced by Gauss's principle.

It is a well-known fact in analytical mechanics that non-holonomic constraints in the form (\ref{eq:Constraints_General}) necessarily require the introduction of Lagrange multipliers in the equations of motion. Only holonomic constraints can be naturally satisfied and, thus, eliminated by using a reduced set of coordinates; i.e., by projecting the dynamics onto the inherent configuration (sub)manifold defined by the holonomic constraints. The pioneering work of Udwadia and Kalaba \cite{pUdwadia_Kalaba_Original} enabled the derivation of equations of motion for mechanical systems subject to constraints in the form  (\ref{eq:Constraints_General}) without the need to introduce Lagrange multipliers. In essence, they provided a means to directly determine the projected dynamics---the dynamics that naturally satisfy the constraints. From the perspective of Gauss's minimization problem, they essentially derived an analytical solution to the problem, yielding a direct expression for the accelerations $\bm{a}$ that is void of $\bm\lambda$.

Exploiting the theory of generalized inverse developed by Moore \citep{Moore_Inverse} and Penrose \citep{Penrose_Inverse}, Udwadia and Kalaba obtained the following analytical solution to the general form of the QP problem (\ref{eq:Gauss_QP_Problem}). If $\bm{q}\in\mathbb{R}^{3N}$ represents an array of generalized coordinates, then Newton's equation of motion can be expressed in terms of $\bm{q}$ as
\begin{equation}\label{eq:Newton_Generalized}
  [\bm{M}(\bm{q})] \ddot{\bm{q}} = \bm{Q}_F + \bm{Q}_R,
\end{equation}
where $\bm{M}$ is a positive definite mass matrix, and $\bm{Q}_F$, $\bm{Q}_R$ are the generalized impressed \footnote{In this formulation, $\bm{Q}_F$ may include inertial (e.g. Coriolis) forces too.} and constraint forces, respectively. In this setting, Gauss's minimization problem is written as
\begin{equation}\label{eq:Gauss_QP_Problem_Generalized}
\begin{array}{llll}

\underset{\ddot{\bm{q}}}{\min} & Z(\ddot{\bm{q}})&=&\frac{1}{2} \left(\ddot{\bm{q}}-\ddot{\bm{q}}^{\rm{free}} \right)^T \bm{M} \left(\ddot{\bm{q}}-\ddot{\bm{q}}^{\rm{free}}\right) \\
\text{s.t.} & [\bm{A}]  \ddot{\bm{q}} &=& \bm{b},
\end{array}
\end{equation}
where $\ddot{\bm{q}}^{\rm{free}}=\bm{M}^{-1}\bm{Q}_F$ is the free motion in the sense of Gauss, which is known---by definition---since the impressed forces are known either directly or through a constitutive law. Udwadia and Kalaba \citep{Udwadia_Kalaba_Original} showed that the ``\textit{optimum}" acceleration is given by
\begin{equation}\label{eq:UK_Solution}
\ddot{\bm{q}}=\ddot{\bm{q}}^{\rm{free}}+ \bm{M}^{-1/2}\left(\bm{A}\bm{M}^{-1/2}\right)^+ \left(\bm{b}-\bm{A}\ddot{\bm{q}}^{\rm{free}}\right),
\end{equation}
where the superscript $+$ indicates a Moore-Penrose inverse \citep{Udwadia_Kalaba_Book}. The strength of Udwadia-Kalaba's solution lies in the fact that Eq. (\ref{eq:UK_Solution}) is ready for direct time integration, as the constraint forces $\bm{Q}_R$ have been eliminated.

\section{The Principle of Minimum Pressure Gradient}
\subsection{Philosophy}
In the last section, we showed the equivalence between Newton's equation and Gauss's principle; fundamentally, Newton's equation represents the first-order necessary condition for minimizing the Gaussian cost. In other words, the \textit{unique} evolution that minimizes the Gaussian cost at each instant must necessarily satisfy Newton's equations of motion, which transforms a dynamics problem into a pure minimization where dynamicists may focus solely on minimizing the Gaussian cost without invoking Newton's equations. The resulting motion is guaranteed to naturally satisfy Newton's equations.

The continuum-mechanics extension of Newton's equation for real fluids is the well-known Navier-Stokes equation:
\begin{equation}\label{eq:NS}
 \rho\bm{a}=-\bm\nabla p + \bm\nabla\cdot\bm\tau + \bm{F},
\end{equation}
where $\bm{a} = \frac{\partial \bm{u}}{\partial t} +\bm{u}\cdot\bm\nabla \bm{u}$ is the inertial acceleration, $\bm\nabla\cdot\bm\tau$ is the impressed viscous force, and $\bm{F}$ is an arbitrarily impressed force (e.g., electromagnetic).

In the realm of analytical and variational mechanics, the Principle of Minimum Pressure Gradient represents the continuum-mechanics extension of Gauss's principle to incompressible fluids \citep{PMPG_PoF}. According to Gauss's principle, the deviation between the actual acceleration $\bm{a}$ and the free acceleration (induced by impressed forces) must be minimum at each instant, provided that the constraint is satisfied. For incompressible flows, the motion is constrained to satisfy the continuity constraint $\bm\nabla\cdot\bm{u}=0$, which is ensured by the pressure force $\bm\nabla p$; i.e., the pressure force acts as the constraint force that ensures the continuity constraint \citep[e.g., ][]{Pressure_BCs,Chorin_Marsden_Book,Pressure_Constraint_General,Morrison2020lagrangian,Evans_PDEs}. Hence, according to Gauss's philosophy, the quantity
\begin{equation}\label{eq:PMPG_Cost}
\mathcal{A}(\bm{a})=\frac{1}{2} \int_\Omega \rho \left[\bm{a}-\frac{1}{\rho}\left(\bm\nabla\cdot\bm\tau + \bm{F} \right)\right]^2 d\bm{x}
\end{equation}
must be minimum at each instant, where $\Omega\subset\mathbb{R}^3$ is the spatial domain.
Note that the cost $\mathcal{A}$ is equivalent to
\[ \mathcal{A}=\frac{1}{2} \int_\Omega \frac{1}{\rho} \left[\bm\nabla p\right]^2 d\bm{x}. \]
That is, similar to Gauss's philosophy, the cost function is a measure of the magnitude of the constraint force required to ensure the constraint.

Since Newton's equation represents the first-order necessary condition for minimizing the Gaussian cost $Z$ in particle mechanics, it is natural to expect that Navier-Stokes' equation (\ref{eq:NS}) serves as the first-order necessary condition for minimizing the pressure gradient cost $\mathcal{A}$ in Eq. (\ref{eq:PMPG_Cost}). This equivalence is established in the theorem below, originally proved in \citep{PMPG_PoF}.\\

\subsection{Mathematical Foundation}
\noindent\textbf{Theorem 1 \citep{PMPG_PoF}} Let $\rho$ be a positive constant. Consider a candidate smooth ($C^\infty$) flow field $\bm{u}(\bm{x};t)$, defined over the domain $\Omega\subset\mathbb{R}^3$ with a smooth boundary $\partial\Omega$ and a time interval $t\in[0,T]$ for some $T>0$ such that: (i) $\bm{u}\cdot\bm\nabla \bm{u}$, $\bm\nabla \cdot \bm\tau$, $\bm{F}$ are in $\mathbb{L}^2(\Omega)$ for all $t\in[0,T]$, and (ii) the initial condition $\bm{u}(\bm{x};0)$ is kinematically-admissible:
\[ \bm\nabla\cdot\bm{u}(\bm{x};0)=0 \; \forall \; \bm{x}\in\Omega \;\; \mbox{and}\;\; \bm{u}(\bm{x};0)\cdot\bm{n}=g(\bm{x}) \; \forall \; \bm{x}\in\partial\Omega, \]
where $\bm{n}$ is the unit normal to the boundary $\partial\Omega$ and $g$ is a given smooth function on $\partial\Omega$.\\

If for every $t\in[0,T]$, the local acceleration $\bm{u}_t(\bm{x};t)\equiv \frac{\partial \bm{u}}{\partial t}(\bm{x};t)$ minimizes the functional
\begin{equation}\label{eq:PMPG_Cost_ut}
 \mathcal{A}(\bm{u}_t) = \frac{1}{2}\int_\Omega \rho \left[\bm{u}_t+\bm{u}\cdot\bm\nabla \bm{u}-\frac{1}{\rho}\left(\bm\nabla\cdot\bm{\tau}+\bm{F}\right)\right]^2 d\bm{x}
 \end{equation}
over the space of admissible \textit{evolutions}:
\[ \Theta=\{\bm{u}_t \in\mathbb{L}^2| \; \bm\nabla \cdot\bm{u}_t=0 \; \mbox{in} \; \Omega, \;\; \bm{u}_t \cdot \bm{n} = 0\; \mbox{on} \; \partial\Omega\}, \]
then the candidate flow field $\bm{u}(\bm{x};t)$ must satisfy the Navier-Stokes equations
\[ \begin{array}{c}
\rho\left(\bm{u}_t+\bm{u}\cdot\bm\nabla \bm{u}\right)=-\bm\nabla p+\bm\nabla \cdot \bm\tau+\bm{F}, \\
\bm\nabla\cdot\bm{u}= 0 \end{array}\]
for all $\bm{x}\in\Omega$, $t\in[0,T]$, for some differentiable function $p$ on $\Omega\times[0,T]$, along with the no-penetration boundary condition
\[ \bm{u}(\bm{x};t)\cdot\bm{n}=g(\bm{x}) \; \forall \; \bm{x}\in\partial\Omega \;\; t\in[0,T]. \]

The proof of this theorem was provided in Ref. \citep{PMPG_PoF}. However, to make this paper self-contained, we reproduce it in Appendix \ref{app:theorem}.

The above theorem implies that Navier-Stokes' equation is the first-order necessary condition for minimizing the pressure gradient cost (\ref{eq:PMPG_Cost}). That is, a candidate solution whose evolution $\bm{u}_t$ minimizes the $\mathbb{L}^2$-norm of the pressure gradient at every instant is guaranteed to naturally satisfy the Navier-Stokes equation. This result transforms an incompressible fluid mechanics problem into a minimization problem, allowing one to focus solely on minimizing the cost $\mathcal{A}$ without invoking the Navier-Stokes equation. This philosophy was recently demonstrated utilizing a neural-network framework in tackling the optimization problem \citep{PMPG_PINN_Daqaq,PMPG_PINN_PoF1,Hussam_Unsteady_ArXiv}. It was also exploited in our recent efforts in developing a theory of lift \citep{Variational_Lift_JFM,Shorbagy_Magnus_AIAA} as well as mathematical modeling of separating flows \citep{Shorbagy_Separation_ArXiv}.

Using standard tools of calculus of variations \citep[e.g., ][]{COV_Dacorogna}, it is straightforward to prove existence and uniqueness of an instantaneous minimizer $\bm{u}_t$ for the functional $\mathcal{A}$---it is basically the minimization problem formulation equivalent to the Leray projection. Additionally, since the Navier-Stokes equation with smooth initial data is known to have a unique solution locally in time (i.e., over $[0,T]$ for some $T>0$), as proved by \cite{Leray_NS}, it follows that the evolution of the Navier-Stokes solution must minimize the pressure gradient cost at each instant. Otherwise, we will have multiple solutions of Navier-Stokes for the same initial and boundary data, which contradicts the proved uniqueness. In other words, a smooth flow field is a solution of the Navier-Stokes equation \textbf{\textit{if and only if}} its evolution minimizes the pressure gradient cost at each instant. In conclusion, an incompressible flow indeed evolves from one instant to the next by minimizing the total magnitude of the pressure gradient force required to ensure the continuity constraint. Any alternative flow candidate would require an unnecessarily larger pressure gradient force to ensure continuity, which contradicts physical considerations as hypothesized by Gauss.

\subsection{Clarifying Common Misunderstandings Surrounding the PMPG}
Because the PMPG is a relatively recent development in the fluid mechanics literature, and because it builds on Gauss's principle of least constraint---a concept not widely emphasized in most engineering curricula---it is understandable that readers may raise questions or express concerns. This section aims to clarify some of the recurring points of confusion about the PMPG.

First, it is important to stress that the above theorem does not imply that the cost functional $\mathcal{A}$ is minimum with respect to the velocity field $\bm{u}(\bm{x};t)$. Indeed, if one sets the first variation of the functional $\mathcal{A}$ with respect to the function $\bm{u}(\bm{x};t)$ to zero, the Navier-Stokes equation will not be obtained as a necessary condition. Rather, Navier-Stokes' equation is the necessary condition for minimizing the pressure gradient functional $\mathcal{A}$ with respect to the local acceleration $\bm{u}_t(\bm{x};t)$ as a function of space, while time appears as a parameter. This approach matches the philosophy of Gauss's principle as discussed in the previous section and demonstrated on a simple example in Ref. \citep{PMPG_PoF}.

Second, despite its name, the \textit{Principle of Minimum Pressure Gradient} does not involve minimizing the functional $\int|\bm\nabla p|^2 d\bm{x}$ over scalar pressure fields. That is, the PMPG is not formulated as a variational principle for $p$ itself. Rather, it seeks the acceleration vector field $\bm{u}_t$ that minimizes the cost functional $\mathcal{A}$, defined in Eq. (\ref{eq:PMPG_Cost_ut}). Although the two cost functionals may appear similar—--both reflecting the $\mathbb{L}^2$-norm of the pressure gradient---they are only equivalent at the minimizing solution. For a general  kinematically-admissible candidate $\bm{u}_t\in\Theta$, the integrand of the cost $\mathcal{A}$ is not necessarily a gradient field; it is guaranteed to be a gradient only at the minimizing $\bm{u}_t$.

Indeed, the two variational formulations are fundamentally different. For example, classical potential theory shows—--via the Dirichlet principle \citep[e.g., ][]{COV_Dacorogna}---that minimizing $\int|\bm\nabla p|^2 d\bm{x}$ (with Dirichlet a boundary condition on $p$) leads to the Laplace equation: $\bm\nabla^2 p=0$. In contrast, Theorem 1 (see its proof in Appendix \ref{app:theorem}) implies that the minimizing solution $\bm{u}_t$ (subject to the continuity constraint $\bm\nabla \cdot\bm{u}_t=0$ and the no-penetration boundary condition $\bm{u}_t\cdot\bm{n}=0$) satisfies the Navier-Stokes equation. Consequently, the corresponding pressure $p$ (identified as the Lagrange multiplier associated with the continuity constraint) satisfies the Poisson equation
\[ \bm\nabla^2 p = \nabla\cdot\bm{a}^{\rm{free}} = \nabla\cdot\left[-\bm{u}\cdot\bm\nabla \bm{u}+\frac{1}{\rho}\left(\bm\nabla\cdot\bm{\tau}+\bm{F}\right)\right].\]
Hence, the two variational formulations differ not only in their variables but also in their boundary conditions and resulting governing equations.

Finally, it may be prudent to emphasize that, while this effort was inspired by Gauss's principle, the above theorem does not rely on it, nor does it assume that the pressure force acts as a constraint force. Rather, the theorem is merely in agreement with these physical concepts. Furthermore, the theorem is broadly applicable---to both laminar and turbulent flows, viscous and inviscid fluids, and Newtonian and non-Newtonian fluids under arbitrary applied forcing $\bm{F}$. The primary requirement is incompressibility.

\section{Fluid Mechanics as a Minimization Problem}
The strength of Gauss's principle lies in its ability to transform mechanics into an optimization problem; indeed, a dynamics problem can be solved purely through optimization using Gauss's principle. In a similar spirit, the PMPG transforms incompressible fluid mechanics into an optimization framework. According to the PMPG, the evolution of an incompressible flow is determined at each instant by minimizing the magnitude of the pressure gradient force required to ensure the continuity constraint. Equivalently, the flow evolves in the closest possible manner to the free motion that would occur in the absence of the continuity constraint. That is, the local acceleration $\bm{u}_t$ must be the closest to the free local acceleration
\begin{equation}\label{eq:Free_Acceleration}
\bm{u}_t^{\rm{free}}(\bm{u}(\bm{x}),\bm{x})=-\bm{u}\cdot\bm\nabla \bm{u}+\frac{1}{\rho}\left( \bm\nabla\cdot\bm{\tau}+\bm{F}\right)
\end{equation}
among all divergence-free candidates satisfying $\bm\nabla\cdot\bm{u}_t=0$, $\bm{u}_t \cdot\bm{n}=0$. \\

Consider a discretized computational domain with an array of nodal values $\bm{U}\in\mathbb{R}^{dN}$, where $d$ is the spatial dimension and $N$ is the total number of mesh points. The cost function $\mathcal{A}$ in Eq. (\ref{eq:PMPG_Cost}) can then be expressed in terms of these nodal values as
\begin{equation}\label{eq:PMPG_Discrete_Cost}
\mathcal{A}(\dot{\bm{U}})=\frac{1}{2} \left(\dot{\bm{U}}-\dot{\bm{U}}^{\rm{free}}(\bm{U}) \right)^T \bm{M} \left(\dot{\bm{U}}-\dot{\bm{U}}^{\rm{free}}(\bm{U})\right),
\end{equation}
where $\dot{\bm{U}}^{\rm{free}}\in\mathbb{R}^{dN}$ is the array of discretized free acceleration $\bm{u}_t^{\rm{free}}$, given in Eq. (\ref{eq:Free_Acceleration}), and $\bm{M}$ is a positive-definite mass matrix that depends on the discretization scheme (it is typically a diagonal matrix in finite difference methods). Additionally, the continuity constraint, imposed on the local acceleration $\bm\nabla\cdot\bm{u}_t=0$, is written in discrete form as:
\begin{equation}\label{eq:Discrete_Continuity}
[\bm{D}]  \dot{\bm{U}}= \bm{0},
\end{equation}
where $\bm{D}\in\mathbb{R}^{N\times dN}$ is the discrete divergence operator---a highly sparse matrix. As such, the PMPG optimization problem is written as
\begin{equation}\label{eq:PMPG_QP_Problem}
\boxed{
\begin{aligned}
\min_{\dot{\bm{U}}} \; \mathcal{A}(\dot{\bm{U}}) &= \frac{1}{2} \left(\dot{\bm{U}} - \dot{\bm{U}}^{\rm{free}} \right)^T \bm{M} \left(\dot{\bm{U}} - \dot{\bm{U}}^{\rm{free}}\right) \\
\text{s.t.} \quad & [\bm{D}] \, \dot{\bm{U}} = \bm{0}
\end{aligned}
}
\end{equation}
which is a convex QP problem since $\bm{M}$ is positive definite. In fact, a structured grid will lead to a strongly positive-definite $\bm{M}$, resulting in a strongly convex QP problem. This formulation is for fixed boundary geometry and stationary boundary conditions, but it can be easily extended to other scenarios for which $\bm{D}$ will be time-varying and the constraint (\ref{eq:Discrete_Continuity}) will be replaced by
\[ [\bm{D}(t)]\dot{\bm{U}}(t)=\bm{b}(t) \]
for some array $\bm{b}$ that includes the instantaneous non-homogenous boundary conditions.

Given an initial condition $\bm{U}(0)$ that is divergence-free ($[\bm{D}]  \bm{U}(0)= \bm{0}$) and satisfies the essential boundary conditions (e.g., no-penetration and no-slip), the free acceleration $\dot{\bm{U}}^{\rm{free}}((\bm{U}(0))$ can be computed---this is precisely the first step in Chorin's projection method \citep{Chorin_Projection,Moin_Incompressible1,Pressure_BCs}. Subsequently, the QP problem (\ref{eq:PMPG_QP_Problem}), formulated under the PMPG framework, can be solved using a standard optimization algorithm to determine the constrained acceleration $\dot{\bm{U}}(0)$, which can then be used to advance the flow forward in time using, for example, the explicit Euler method:
\[ \bm{U}(k+1) = \bm{U}(k) + \Delta t \dot{\bm{U}}(k) \]
for a suitably chosen time step $\Delta t$.

The procedure described above avoids the most computationally demanding step in projection methods---solving the Poisson equation in pressure \citep{Poisson_Equation_Pressure_Cost}, which accounts for 40\% to 90\% of the total computational cost, depending on the problem and computational setup. A fundamental challenge in standard simulation techniques for the incompressible Navier-Stokes equation is the absence of an update equation for pressure (i.e., there is no explicit equation for $\frac{\partial p}{\partial t}$). Specifically, if the velocity field $\bm{u}(\bm{x},t)$ is known at some time instant $t$, the momentum equation (Navier-Stokes) can be used to directly update the velocity field at the next time step, $\bm{u}(\bm{x},t+\Delta t)$, provided that the pressure field $p$ is known. However, no analogous equation exists for updating the pressure field; instead, pressure is implicitly determined through the continuity constraint. Several techniques have been developed to resolve this issue and allow for simultaneous update of the pressure and velocity fields without an explicit evolution equation for the former. The two most common approaches are: (i) Iterative methods, such as the Semi-Implicit Method for Pressure-Linked Equations (SIMPLE) by Patankar and Spalding \citep{Patanker_Spalding_SIMPLE}, widely adopted in commercial computational fluid dynamics (CFD) software (e.g., ANSYS); and (ii) Projection-based methods, such as the Chorin-Temam projection technique \citep{Chorin_Projection,Temam_Projection}. In iterative methods, the pressure field is initially guessed, allowing computation of the velocity field from the momentum equation. The Poisson equation is then solved to obtain a pressure correction. In the second approach, the Navier-Stokes equation is projected onto the space of divergence-free fields by solving the Poisson equation in pressure at each time step. The updated pressure field is then used to correct the velocity field \footnote{Another alternative is the artificial compressibility approach \citep{Chorin_Artificial_Compressibility}.}. The extensive literature on this topic extends far beyond the scope of this paper \citep[e.g.,][]{Projection_Book,Projection_Review,Hirsch_Book2}.

In both approaches, solving the Poisson equation at each time step constitutes a significant computational bottleneck, posing a major obstacle to high-fidelity simulations of large-scale turbulent flow fields. A potential strength of the PMPG formulation lies in eliminating the need to solve a Poisson equation at every time step. Instead, it replaces this task with a convex QP problem, which is perhaps the most efficient class of problems in non-linear optimization, with numerous well-established algorithms available. Furthermore, the convex QP problem at hand has a highly favorable structure; it has a diagonal Hessian $\bm{M}$ and the
constraint matrix $\bm{D}$ is extremely sparse---each row has only a few non-zero entries irrespective of the mesh resolution. Nevertheless, it is not clear whether the presented convex QP formulation of fluid mechanics will outperform the state-of-the-art projection methods (e.g., multigrid methods \citep{Projection_Review}), which scales as $N \log N$. A detailed comparison between the QP formulation and standard projection methods in terms of computational cost is beyond the scope of this paper and will be addressed in future work. Here, we focus solely on formulating the fluid mechanics problem into a convex QP framework. Furthermore, in the next section, we derive an analytical solution to this QP problem, providing an explicit ODE for the projected dynamics of Navier-Stokes, ready for direct time integration.

\section{Variational Projection of Navier-Stokes for Direct Time Integration}
It should be noted that casting an incompressible fluid mechanics problem within a minimization framework might not be seen as novel in the eyes of certain experts in computational fluid dynamics. While rarely stated explicitly, the idea has been implicitly referenced in a few efforts. For instance, in their quantum-inspired study of turbulence structure, Gourianov et al. \citep{Nature_Communication_Minimization} implemented a numerical simulation scheme for the incompressible Navier-Stokes equation that minimizes the following quantity at each time step (see Eq. (8) in their reference):
\begin{equation}\label{eq:Nature_CS_Formulation}
\left|\left|\frac{\bm{V}^*-\bm{V}}{\Delta t}+\bm{V}\cdot\bm\nabla \bm{V} - \nu\bm\nabla^2 \bm{V}\right|\right|_2^2 + \mu ||\bm\nabla \cdot \bm{V}^*||_2^2,
\end{equation}
where $||.||_2$ denotes the $L^2$-norm, and $\mu$ is a weighting factor. Notably, the first term is precisely our cost functional $\mathcal{A}$---the $L^2$-norm of the pressure gradient---for Newtonian fluids when using first-order Euler time integration. The second term is simply imposing the continuity constraint on the updated velocity $\bm{V}^*$ using the penalty method, which augments the cost with a positive term that increases as the constraint violation grows \citep{Penalty_Method,Penalty_ALM}.

Aside from the fact that the penalty method is far from being the most suitable approach for enforcing the incompressibility constraint, this formulation is, in principle, equivalent to the QP formulation presented in the previous section. However, the authors of \citep{Nature_Communication_Minimization} have not justified their minimization framework in contrast to our QP formulation, which is rigorously justified by the PMPG theorem and deeply rooted in analytical mechanics via Gauss's principle. Indeed, without the PMPG, there is no fundamental guarantee that the minimizing solution from Eq. (\ref{eq:Nature_CS_Formulation}) satisfies the Navier-Stokes equation. Theorem 1 explicitly establishes this link. One may argue, however, that the minimization formulation in Eq. (\ref{eq:Nature_CS_Formulation}) is analogous to Chorin's projection, which is well-justified---conceptually, any projection problem can be casted into a minimization problem. Nevertheless, proper formulation following concepts from analytical mechanics such as Gauss's principle and the PMPG provides key advantages: (i) It identifies the minimization formulation as a QP problem; and (ii) It enables the application of analytical-mechanics tools of Udwadia and Kalaba \citep{Udwadia_Kalaba_Book} to derive an explicit analytical solution for the problem---a task that has remained elusive for decades, likely due to the absence of a proper analytical-mechanics formulation.

The QP formulation, presented in Eq. (\ref{eq:PMPG_QP_Problem}), can be solved analytically using the Moore-Penrose theory of generalized inverse \citep{Moore_Inverse,Penrose_Inverse}, following Udwadia and Kalaba pioneering efforts in analytical mechanics \citep{Udwadia_Kalaba_Original}. For simplicity, we will follow Udwadia-Kalaba's scaling:
\[ \tilde{\bm{U}} = \bm{M}^{1/2}\; \bm{U}, \;\;  \tilde{\bm{D}} = \bm{M}^{-1/2} \; \bm{D}. \]
In this transformed coordinate-system, the QP problem (\ref{eq:PMPG_QP_Problem}) is written as
\begin{equation}\label{eq:PMPG_QP_Problem_Scaling}
\begin{aligned}
\min_{\dot{\tilde{\bm{U}}}} \; \mathcal{A}(\dot{\tilde{\bm{U}}}) 
&= \frac{1}{2} \left(\dot{\tilde{\bm{U}}} - \dot{\tilde{\bm{U}}}^{\rm{free}} \right)^T 
         \left(\dot{\tilde{\bm{U}}} - \dot{\tilde{\bm{U}}}^{\rm{free}} \right) \\
\text{s.t.} \quad & [\tilde{\bm{D}}] \, \dot{\tilde{\bm{U}}} = \bm{0}
\end{aligned}
\end{equation}

For an under-determined linear system
\[ \bm{A}\bm{x}=\bm{b},\;\;  \bm{x}\in\mathbb{R}^n, \;\; \bm{b}\in\mathbb{R}^m, \;\; \mbox{and} \;\; m<n, \]
the theory of generalized inverse implies that the solution $\bm{x}$ is written as
\begin{equation}\label{eq:Generalized_Inverse}
\bm{x}=\bm{A}^+\bm{b} + \left(\bm{I}-\bm{A}^+\bm{A}\right)\bm{y},
\end{equation}
where $\bm{A}^+$ is the Moore-Penrose inverse of $\bm{A}$, and $\bm{y}$ is arbitrary. The matrix $\bm{A}^+\bm{A}$ is a symmetric projection matrix; i.e., $\left(\bm{A}^+\bm{A}\right)^2=\bm{A}^+\bm{A}$. As such, the solution $\dot{\tilde{\bm{U}}}$ to the linear constraint: $\tilde{\bm{D}} \dot{\tilde{\bm{U}}}= \bm{0}$ can be written as
\begin{equation}\label{eq:Generalized_Inverse_Divergence}
\dot{\tilde{\bm{U}}} = \bm{N} \bm{y}, \;\; \bm{N}=\bm{I} - \bm{P}_D, \;\; \bm{P}_D=\tilde{\bm{D}}^+ \tilde{\bm{D}}
\end{equation}
where $\bm{y}$ is arbitrary, $\tilde{\bm{D}}^+$ is the Moore-Penrose inverse of $\tilde{\bm{D}}$. The matrices $\bm{P}_D$, $\bm{N}$ are symmetric projection matrices; the former projects on the range space of $\tilde{\bm{D}}$ and the latter projects on the null space of $\tilde{\bm{D}}$. Therefore, given a vector field $\bm{v}$ whose scaled discretization is represented by the array $\tilde{\bm{V}}$, the Helmholtz-Hodge decomposition \citep{Helmholtz_Decomposition_Review} into divergence-free and curl-free components is simply written as
\[ \tilde{\bm{V}} = \underbrace{[\bm{P}_D]\tilde{\bm{V}}}_{\rm{Curl-Free}} + \underbrace{[\bm{N}]\tilde{\bm{V}}}_{\rm{Divergence-Free}}.\]
That is the matrix $\bm{N}$ projects $\tilde{\bm{V}}$ on the space of divergence-free vector fields.

Substituting by $\dot{\tilde{\bm{U}}}$ from Eq. (\ref{eq:Generalized_Inverse_Divergence}) into the cost $\mathcal{A}$ in Eq. (\ref{eq:PMPG_QP_Problem_Scaling}) transforms the problem from a constrained minimization problem into an unconstrained one:
\[ \min_{\bm{y}} \; \mathcal{A}(\dot{\tilde{\bm{U}}}(y))=\frac{1}{2} \left(\bm{N} \bm{y}-\dot{\tilde{\bm{U}}}^{\rm{free}} \right)^T \left(\bm{N} \bm{y}-\dot{\tilde{\bm{U}}}^{\rm{free}}\right) \]
over the free variable $\bm{y}$ since the constraint is automatically satisfied for any $\bm{y}$. Consequently, differentiating $\mathcal{A}$ with respect to the free variable $\bm{y}$, we obtain the first-order necessary condition:
\[ \bm{N}^T\left(\bm{N} \bm{y}-\dot{\tilde{\bm{U}}}^{\rm{free}} \right) = \bm{0} \;\;\Rightarrow\;\; \bm{N} \left(\bm{y}-\dot{\tilde{\bm{U}}}^{\rm{free}}\right)= \bm{0}, \]
which---after applying the generalized inverse formula (\ref{eq:Generalized_Inverse}) again---yields
\[ \bm{y}=\dot{\tilde{\bm{U}}}^{\rm{free}} + \left(\bm{I}-\bm{N}^+\bm{N}\right)\bm{z}, \]
where $\bm{z}$ is arbitrary. Exploiting the properties of Moore-Penrose inverses, one can show (see Eq. (2.79), pp. 50 in \citep{Udwadia_Kalaba_Book}):
\[ \bm{I}-\bm{N}^+\bm{N} = \bm{P}_D. \]
Hence, we have
\[\bm{y}=\dot{\tilde{\bm{U}}}^{\rm{free}} + \bm{P}_D\bm{z}. \]
Substituting by $\bm{y}$ into Eq. (\ref{eq:Generalized_Inverse_Divergence}) and realizing that $\bm{N}\bm{P}_D=\bm{0}$, we obtain
\begin{equation}\label{eq:Projected_NS}
\boxed{\dot{\tilde{\bm{U}}} = [\bm{N}] \dot{\tilde{\bm{U}}}^{\rm{free}}(\tilde{\bm{U}})}.
\end{equation}

Equation (\ref{eq:Projected_NS}) represents the projected dynamics of the spatially-discretized Navier-Stokes equation. We refer to it as Variational Projection of Navier-Stokes (VPNS). This ordinary differential equation (ODE) is ready for direct time integration, eliminating the need for iterative methods (such as the SIMPLE algorithm) or solving the Poisson equation for pressure at each time step. Starting from an initial condition that is divergence-free and satisfies the essential boundary conditions, the evolution under the VPNS guarantees a divergence-free velocity field at all future time steps. To the best of our knowledge, the VPNS formulation in Eq. (\ref{eq:Projected_NS}) is novel and has not been previously derived. Unlike all existing methods for pressure-velocity coupling in incompressible flow, the VPNS provides an explicit formula for the projected dynamics of Navier-Stokes. Notably, a similar abstract formula is well-known:
\[ \bm{u}_t  = \mathcal{N} \bm{a}^{\rm{free}}, \]
where $\mathcal{N}$ is an abstract projection operator (Leray projection) whose application to a given vector field requires solving a partial differential equation---the Poisson equation. In contrast, in the VPNS formulation, projection is performed after spatial discretization, transforming the problem into a linear projection in finite dimensions, which allows exploitation of linear algebraic tools (such as the Moore-Penrose theory of generalized inverses). Although our derivation was not simply a linear projection in finite dimensions, the final result ultimately takes the same form. Consequently, the matrix $\bm{N}$ in the VPNS, as given in Eq. (\ref{eq:Projected_NS}), has an explicit formula in terms of the discrete divergence operator $\bm{D}$:
\[ \bm{N} = \bm{I}-\tilde{\bm{D}}^+ \tilde{\bm{D}}, \;\; \tilde{\bm{D}} = \bm{M}^{-1/2} \; \bm{D}, \]
where $\bm{M}$ is the mass matrix, which is typically diagonal for finite-difference and finite-volume discretizations.

Equation (\ref{eq:Projected_NS}) of the VPNS can be expressed as
\begin{equation}\label{eq:Projected_NS_Detailed}
\dot{\tilde{\bm{U}}} = \underbrace{\dot{\tilde{\bm{U}}}^{\rm{free}}}_{\rm{Prediction}} - \underbrace{\bm{P}_D \dot{\tilde{\bm{U}}}^{\rm{free}}}_{\rm{Correction}}. \end{equation}
The projected dynamics $\dot{\tilde{\bm{U}}}$ consists of two parts: (i) the free acceleration $\dot{\tilde{\bm{U}}}^{\rm{free}}$, which is obtained by ignoring the continuity constraint and the pressure force required to enforce it; and (ii) a correction term, which lies in the range of $\tilde{\bm{D}}$ (i.e., a curl-free component orthogonal to the space of divergence-free vector fields). These two parts correspond precisely to the predictor and corrector steps in Chorin's projection method \cite{Chorin_Projection}. The predictor step is equivalent to determining the free acceleration (\ref{eq:Free_Acceleration}) in the language of Gauss. The correction term ensures incompressibility by subtracting the curl-free component from the free acceleration, leaving only the divergence-free component. The strength of the VPNS formulation lies in providing an explicit formula for this correction term whose computation traditionally requires solving a Poisson equation. This term corresponds exactly to the pressure gradient force: the discretized array $\bm{Q}_R$ of $\bm\nabla p$ (after scaling back) is simply given by:
\[ \bm{Q}_R=[\bm{M}^{-1/2}] [\bm{P}_D] \dot{\tilde{\bm{U}}}^{\rm{free}}. \]

The VPNS formulation in Eq. (\ref{eq:Projected_NS}) is expected to be highly valuable to fluid mechanicians for both simulation and theoretical analysis. Notably, for problems with fixed boundary conditions, the projection matrix $\bm{N}$ is constant; it is computed only once before time marching. Additionally, $\dot{\tilde{\bm{U}}}^{\rm{free}}$ is given in Eq. (\ref{eq:Free_Acceleration}) in terms of the convective acceleration, the viscous stress, and the externally applied force $\bm{F}$, which are known in terms of the instantaneous velocity field $\tilde{\bm{U}}$. For instance, $\dot{\tilde{\bm{U}}}^{\rm{free}}$ is quadratic in $\tilde{\bm{U}}$ for Newtonian fluids. Subsequently, numerical time integration of Eq. (\ref{eq:Projected_NS}) can be implemented directly---it is an explicit nonlinear ODE with constant coefficients. Hence, it can be marched forward using standard time-marching techniques---whether explicit
\[ \tilde{\bm{U}}(k+1) = \tilde{\bm{U}}(k) + \Delta t [\bm{N}] \dot{\tilde{\bm{U}}}^{\rm{free}}(\tilde{\bm{U}}(k)), \]
or implicit
\[ \tilde{\bm{U}}(k+1) = \tilde{\bm{U}}(k) + \Delta t [\bm{N}] \dot{\tilde{\bm{U}}}^{\rm{free}}(\tilde{\bm{U}}(k+1)) \]
This compact form will enable exploitation of the rich legacy of nonlinear systems theory \citep[e.g.,][]{Khalil,Sastry_Book,Bullo_Lewis_Book} in stability and bifurcation analysis as well as control design.

Finally, we find it necessary to clarify that the VPNS approach does not eliminate the global (elliptic) nature of enforcing incompressibility. While it avoides solving the Poisson equation at each time step, it introduces analogous challenges through the computation of the Moore-Penrose inverse $\tilde{\bm{D}}^+$. In principle, the Poisson equation could be solved by precomputing the inverse $\Delta^{-1}$ of the discretized Laplacian, but it is typically avoided due to storage costs and numerical instability. In contrast to $\Delta^{-1}$, the Moore–Penrose inverse $\tilde{\bm{D}}^+$  is not a full inverse of a square matrix, but rather a generalized inverse used solely to project onto the null space of $\tilde{\bm{D}}$. It can be efficiently approximated using numerically-stable, structure-exploiting methods. In this sense, the VPNS replaces solving a sequence of repeated linear systems with a one-time construction of an explicit projection operator. In conclusion, the relative efficiency of VPNS versus standard projection methods (e.g. multigrid solvers) remains an open question and warrants a dedicated quantitative comparison. Regardless of this tradeoff, the VPNS offers a conceptually elegant and variationally grounded formulation that is significantly simpler to implement and analyze.


\section{Demonstration on the Lid-Driven Cavity Problem}
In this section, we demonstrate the VPNS formulation on one of the benchmark problems in fluid mechanics: the unsteady, viscous, incompressible flow in a square cavity whose lid is driven at a constant speed $U_{lid}$. In Appendix \ref{appA}, we provide a detailed construction of the matrices $\bm{M}$ and $\bm{D}$ as well as the array $\dot{\bm{U}}^{\rm{free}}$ of free accelerations, which collectively define the QP problem (\ref{eq:PMPG_QP_Problem}). They also determine the matrices $\tilde{\bm{D}}^+$, $\bm{P}_D$, and $\bm{N}$ and the array $\dot{\tilde{\bm{U}}}^{\rm{free}}$, which are required for the VPNS formulation. Accordingly, we perform direct time integration of the VPNS ODE (\ref{eq:Projected_NS}) and compare the resulting flow field against simulations obtained from OpenFOAM.

\subsection{Validation}
We consider the case of a lid-driven cavity with a Reynolds number of 20 based on the lid velocity $U_{lid}$ and the square cavity dimension $L$. Additionally, we have a non-dimensional time $\tau=\frac{t}{T_{ref}}$ based on the reference time $T_{ref}=\frac{L}{U_{lid}}=50$; i.e., a particle sticking to the lid and traveling with it, if starts at the upper left corner, it will reach the upper right corner in 50 seconds. We perform time marching simulation of the VPNS equation (\ref{eq:Projected_NS}) from a stagnant initial condition $\bm{U}(0)=\bm{0}$ using a simple explicit Euler scheme with time step of $\Delta \tau=8e^{-5}$ for $T=1250$ steps (i.e., over a period of 5 seconds---equivalently $\tau=0.1$). We perform these simulations using four different uniform meshes of $50\times50$, $75\times75$, $100\times100$, and $125\times125$ to study mesh convergence.

\begin{figure*}
\begin{center}
$\begin{array}{cc}
\subfigure[Variation of $u$ with $y$.]{\label{fig:Validationu}\includegraphics[width=6.5cm]{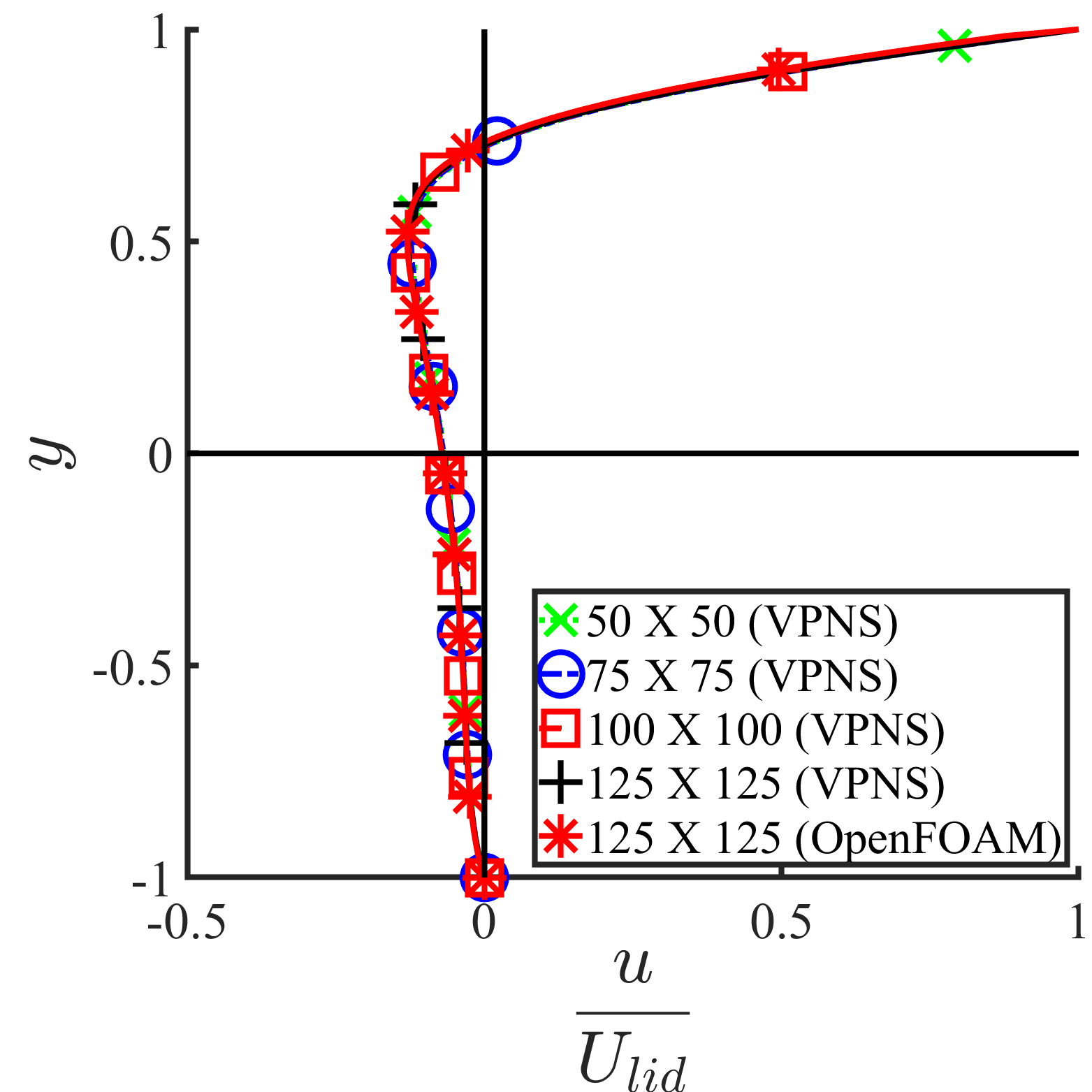}} & \subfigure[Variation of $v$ with $x$.]{\label{fig:Validationv}\includegraphics[width=6.5cm]{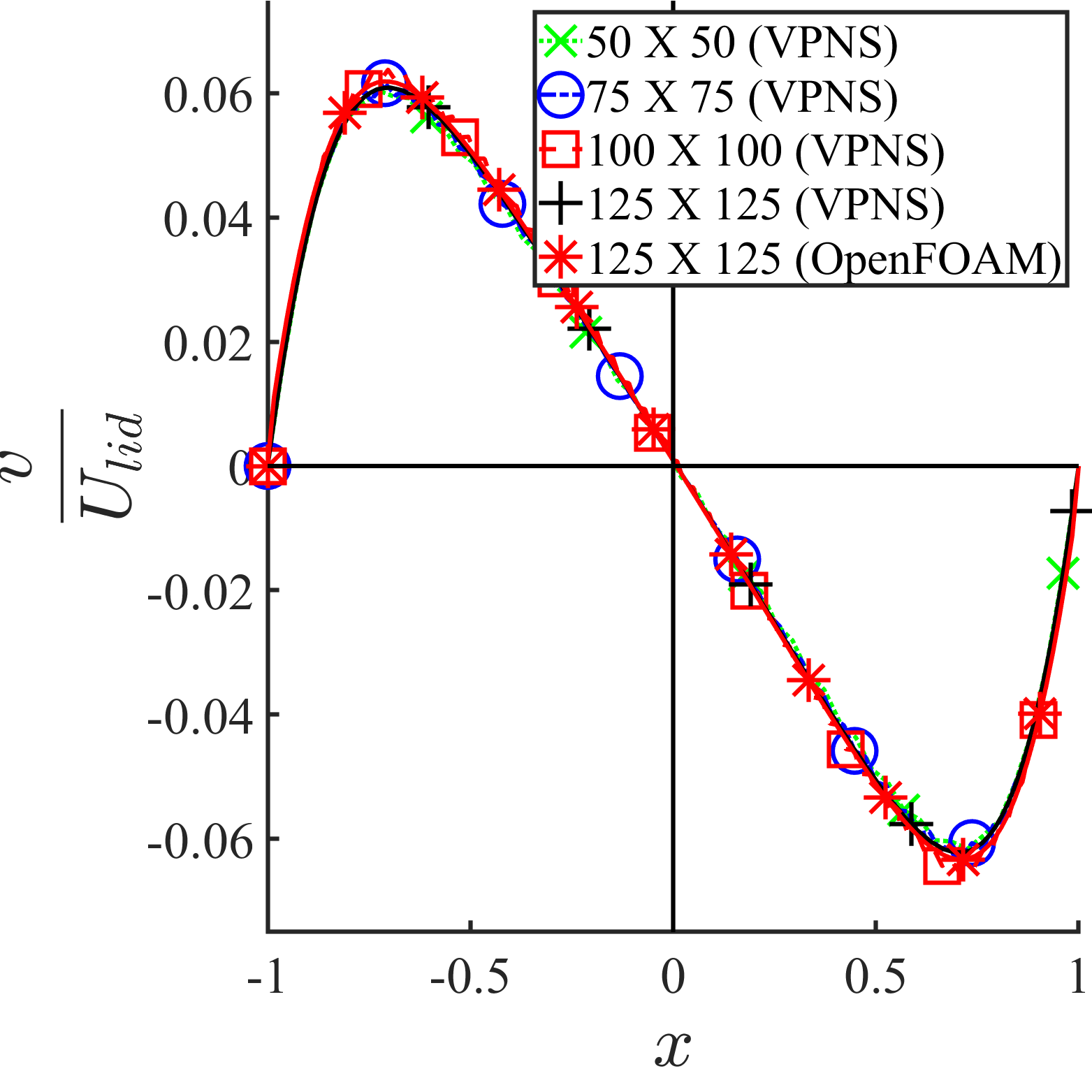}}
\end{array}$
\caption{Comparison between the VPNS simulation over four meshes, and OpenFoam using $125\times125$ in terms of the flow field $u$, $v$ at the final time $T$ along the vertical and horizontal lines, respectively, passing through the mid point. The comparison shows excellent mesh convergence and matching with OpenFOAM.}
\label{Fig:Validation}
\end{center}
\end{figure*}

\begin{figure*}
\begin{center}
\includegraphics[width=18cm]{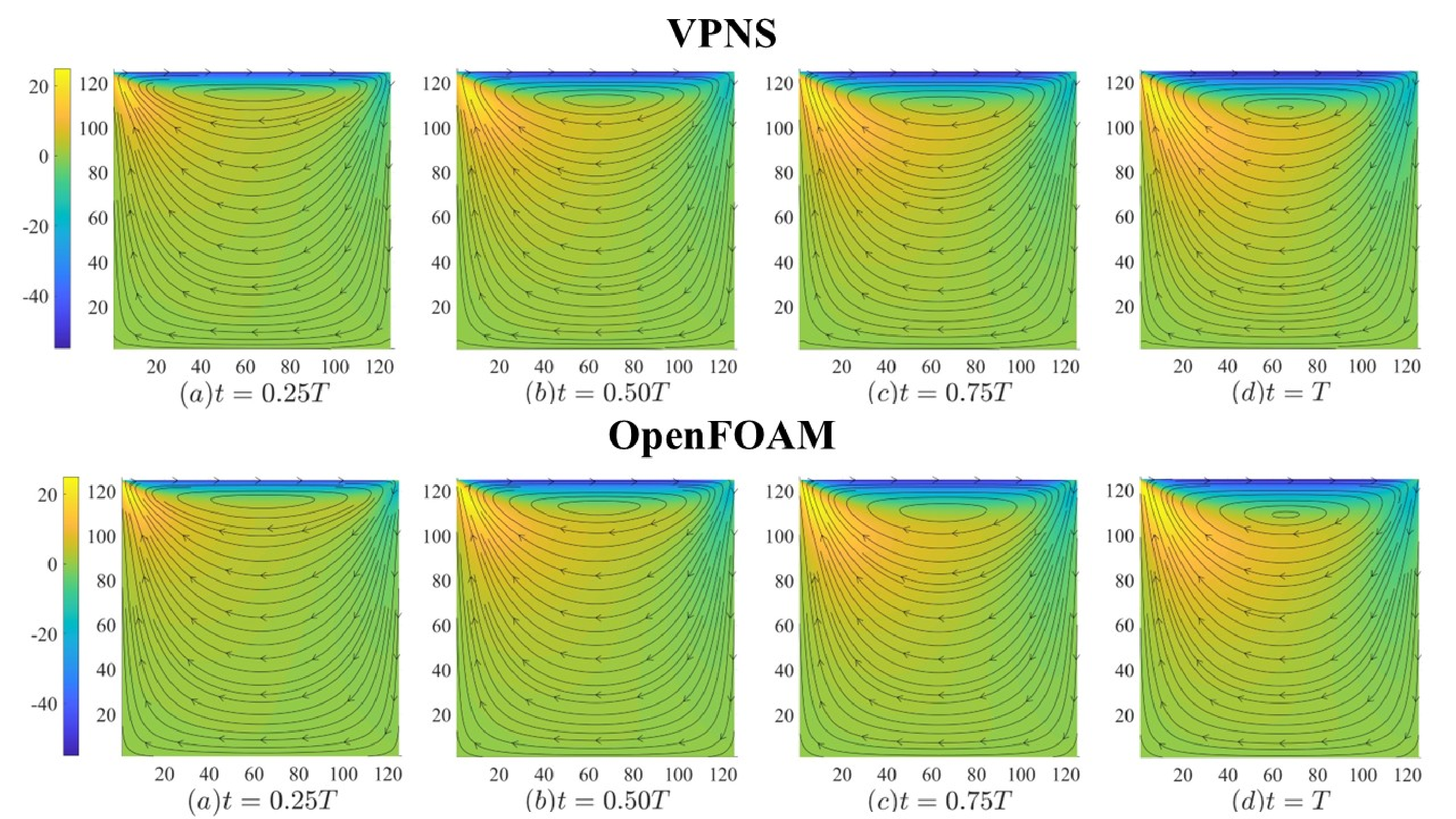}
\caption{Contours of the nondimensional vorticity $\omega T_{ref}$ at $T/4$, $T/2$, $3T/4$, and $T$ from the VPNS and OpenFOAM simulations.}
\label{Fig:Validation_Vorticity}
\end{center}
\end{figure*}

Figure \ref{Fig:Validation} shows both mesh convergence and validation of the simulated flow field against OpenFOAM. The figure shows variations of the horizontal component $u$ (at the final time $T$) with the vertical line passing through the mid point, and the vertical component $v$ (at $T$) with the horizontal line passing through the same point. Results from the VPNS are presented for the four different meshes, mentioned above, demonstrating excellent mesh convergence. The figure also presents the same results from OpenFOAM simulations using the finest mesh of $125\times125$ and the same time step $\Delta t$, which matches the VPNS flow field. Figure \ref{Fig:Validation_Vorticity} shows contours of the nondimensional vorticity $\omega \tau$ at the time instants of $T/4$, $T/2$, $3T/4$, and $T$, from the VPNS and OpenFOAM simulations---demonstrating qualitative matching.

\begin{figure}
 \begin{center}
 \includegraphics[width=6.5cm]{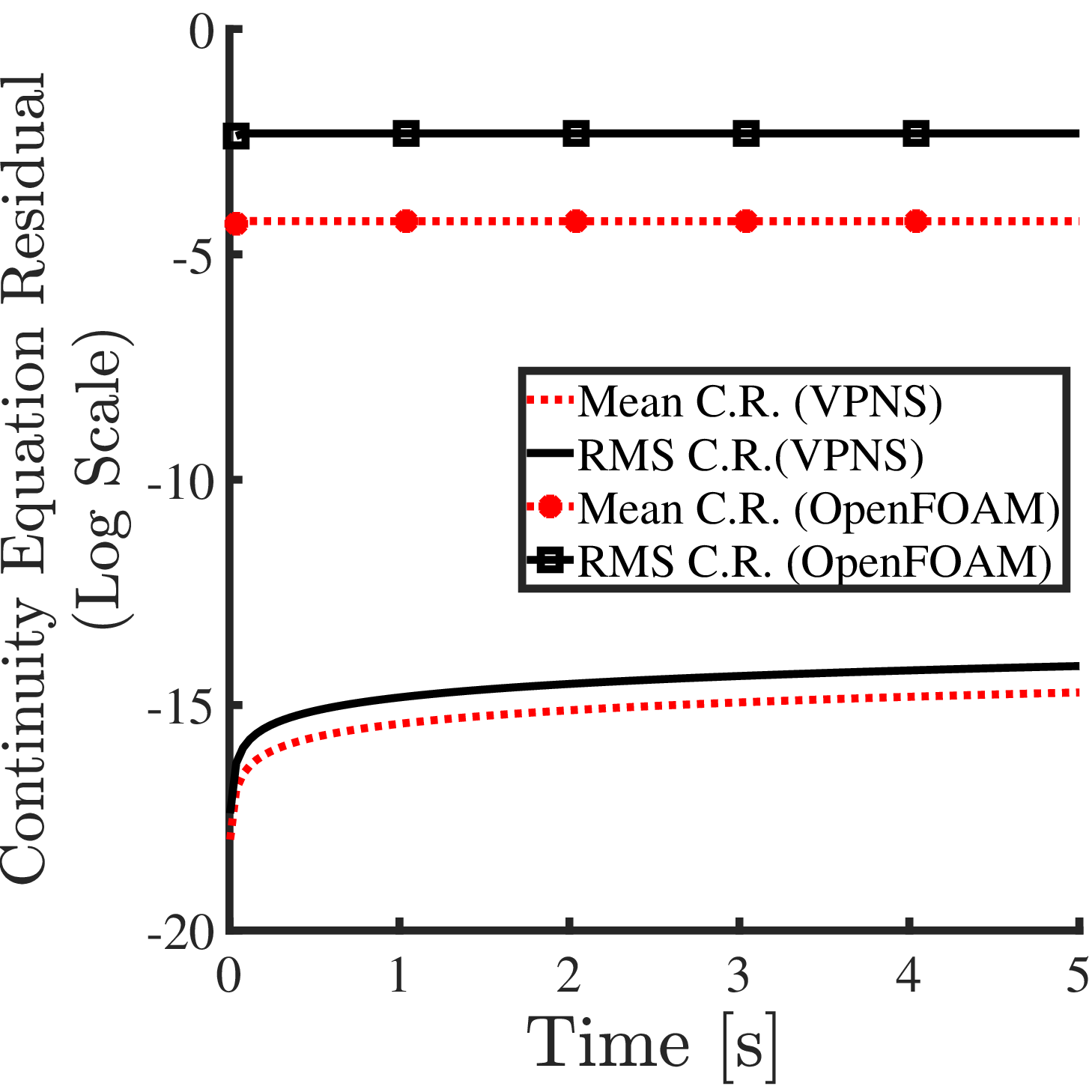}
 \caption{Comparison between the VPNS and OpenFOAM in terms of the RMS and mean continuity residuals. The VPNS formulation is far more accurate in enforcing the continuity constraint.}
 \label{Fig:Continuity}
 \end{center}
\end{figure}

\noindent Figure \ref{Fig:Continuity} shows a comparison between the VPNS and OpenFOAM results in terms of the average and root-mean-square (RMS) residual of the continuity equation, which quantifies violation of the incompressibility constraint. The figure clearly demonstrates that the VPNS formulation enforces the continuity constraint with accuracy several orders of magnitude higher than that of OpenFOAM. In fact, the maximum violation (i.e., maximum of $|\bm\nabla\cdot \bm{u}|$) over the field for the VPNS solution is on the order of $10^{-13}$, whereas OpenFOAM exhibits violations of order 1 at the points adjacent to the lid corners. This remarkable level of precision in enforcing incompressibility is one of the main strengths of the VPNS formulation.\\

\subsection{Demonstration of the Optimal Evolution}
The philosophy of the Principle of Minimum Pressure Gradient (PMPG) implies that, at any given instant, the flow evolves in a way that minimizes the pressure gradient cost $\mathcal{A}$. Among all kinematically admissible evolutions $\dot{\bm{U}}$, Nature picks the one, $\dot{\bm{U}}^*$, that requires the least pressure gradient force to ensure continuity. Any alternative evolution $\dot{\bm{U}}\neq\dot{\bm{U}}^*$ would require an unnecessarily larger pressure gradient force to maintain the continuity constraint.


\begin{figure}
 \begin{center}
 \includegraphics[width=6.5cm]{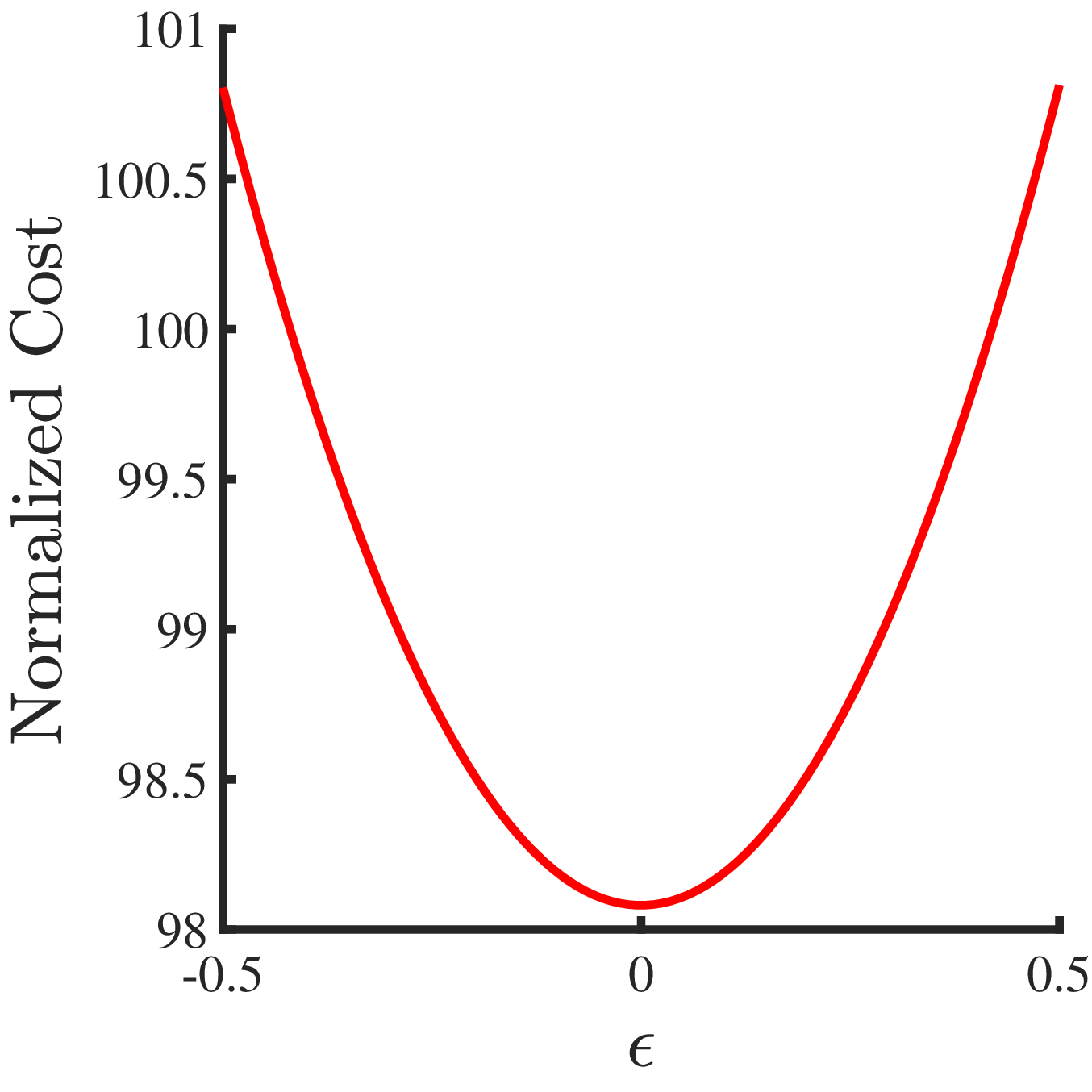}
 \caption{Variation of the normalized cost $\hat{\mathcal{A}}=\frac{\mathcal{A}}{\rho U_{lid}^4}$ with the size $\epsilon$ of perturbation from the true evolution  $\dot{\bm{U}}^*$. The pressure gradient cost attaints its minimum precisely at $\epsilon=0$. That is, the VPNS evolution $\dot{\bm{U}}^*$ minimizes the cost $\mathcal{A}$ over all kinematically admissible evolutions $\dot{\bm{U}} = \dot{\bm{U}}^* + \epsilon \bm\eta$.}
 \label{Fig:Minmiality_Demo1D}
 \end{center}
\end{figure}

This philosophy is manifested in Fig. \ref{Fig:Minmiality_Demo1D}. We select the instant $T/2$ as a representative example for demonstration purposes, without loss of generality. At this instant, the velocity field $\bm{U}$ and the true (optimum) evolution $\dot{\bm{U}}^*$ are available from the VPNS simulation. Consider any kinematically admissible perturbation $\bm\eta$; i.e., divergence-free $[\bm{D}]\bm\eta=0$ and vanishes at the boundaries. We construct a family of legitimate (kinematically admissible) evolutions:
\[ \dot{\bm{U}} = \dot{\bm{U}}^* + \epsilon \bm\eta, \]
where $\epsilon$ is the size of perturbation from the true evolution $\dot{\bm{U}}^*$ in the direction of $\bm\eta$. Since each member of this family satisfies the continuity constraint and boundary conditions, they represent legitimate alternative evolutions. Recall that the pressure gradient cost $\mathcal{A}(t)$ at a given instant depends on the instantanoues velocity field $\bm{U}$ and the evolution $\dot{\bm{U}}$, as given in Eq. (\ref{eq:PMPG_Discrete_Cost}). We then compute $\mathcal{A}(\dot{\bm{U}},\bm{U})$ corresponding to each member in this family of legitimate evolutions. Figure \ref{Fig:Minmiality_Demo1D} shows the variation of the normalized pressure gradient cost $\hat{\mathcal{A}}=\frac{\mathcal{A}}{\rho U_{lid}^4}$ with $\epsilon$. As expected, $\hat{\mathcal{A}}$ attains its minimum precisely at $\epsilon=0$, corresponding to the true evolution $\dot{\bm{U}}^*$, while any deviation results in a higher cost.

\begin{figure*}
\begin{center}
$\begin{array}{cc}
\subfigure[Nondimensional $\bm\eta_1$.]{\label{fig:var_Udot_per1} \includegraphics[width=6.5cm]{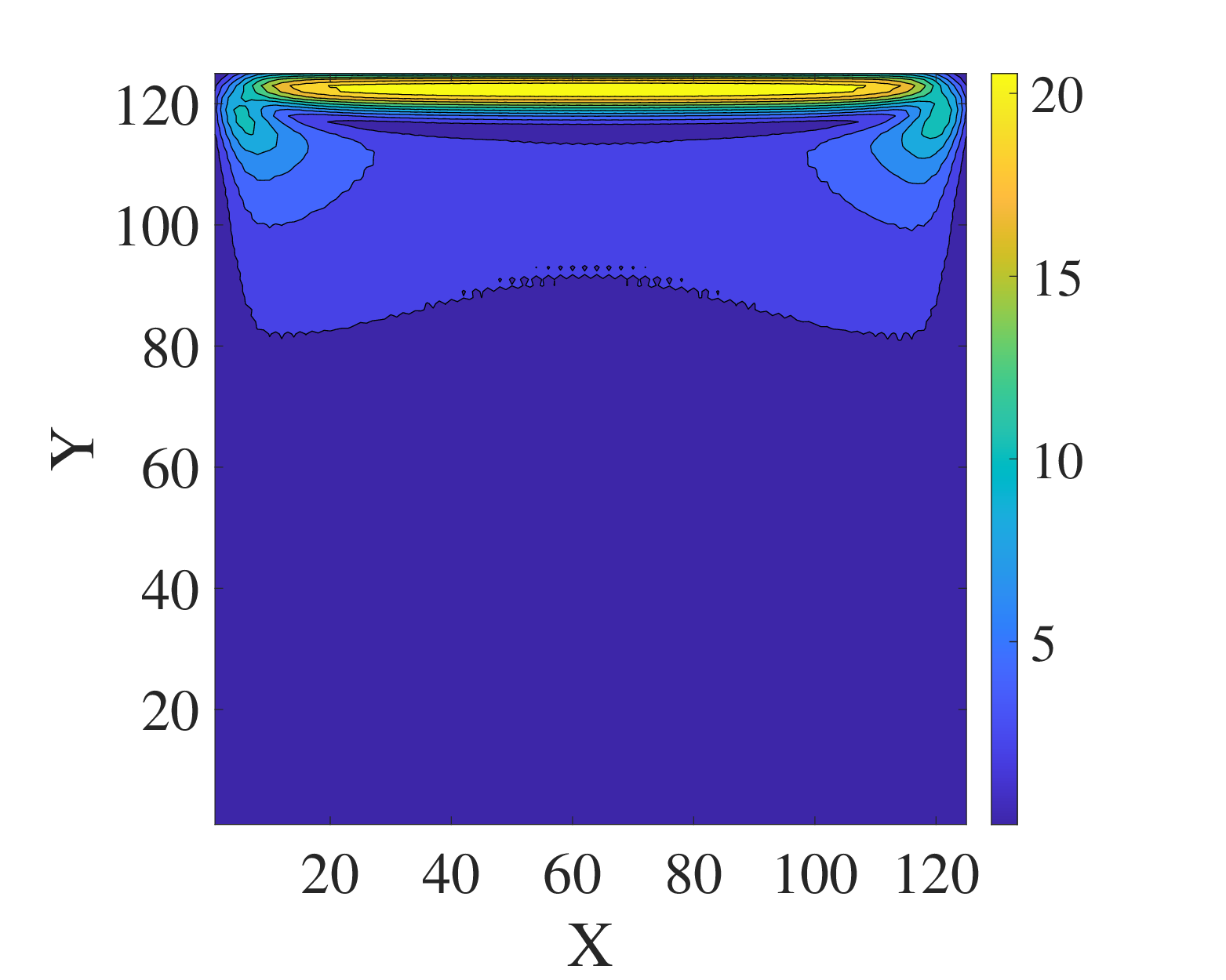}}
& \subfigure[Nondimensional $\bm\eta_2$.]{\label{fig:var_Udot_per2}\includegraphics[width=6.5cm]{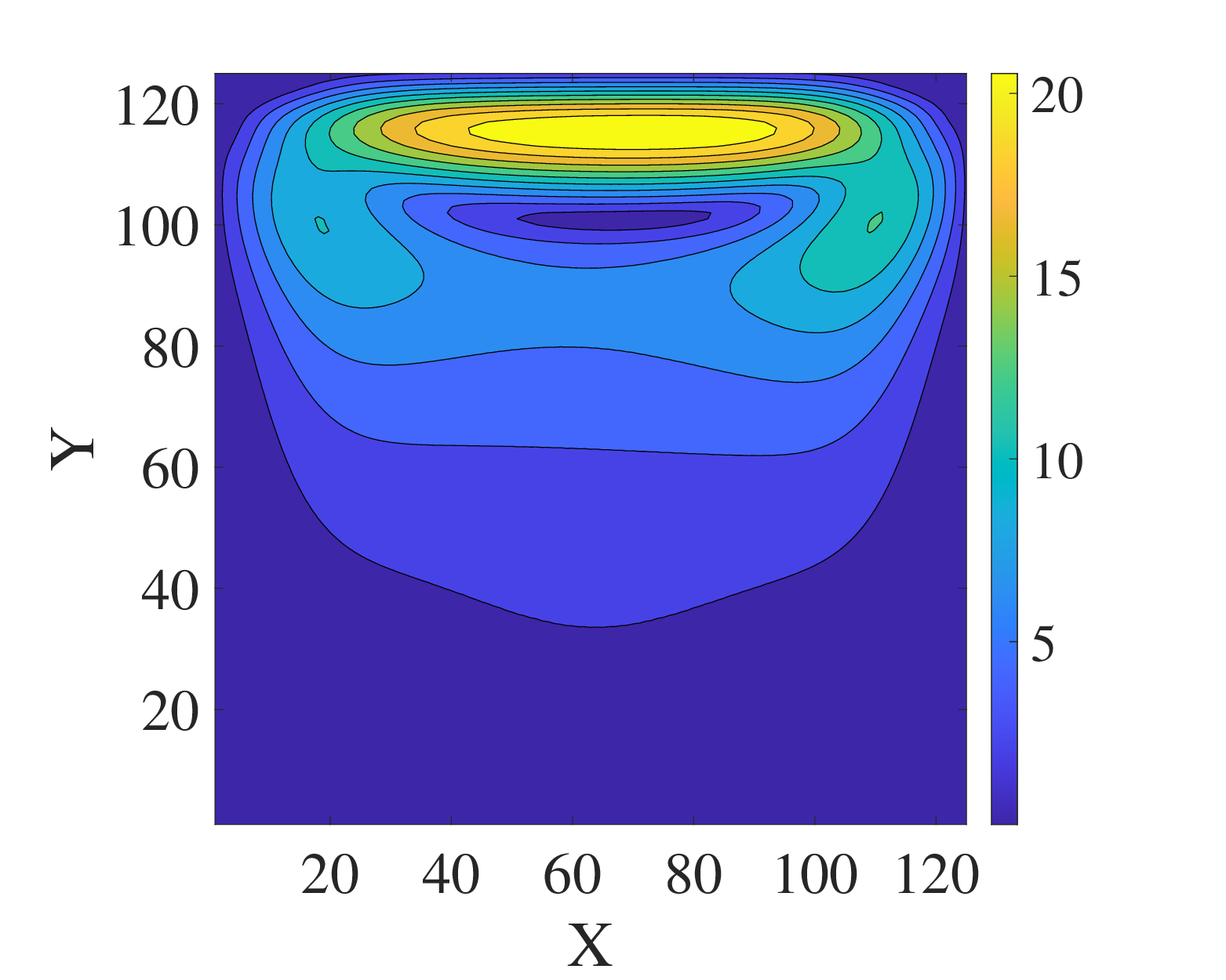}}\\

\subfigure[Nondimensional $\dot{\bm{U}}^*$.]{\label{fig:var_Udot_tby2} \includegraphics[width=6.5cm]{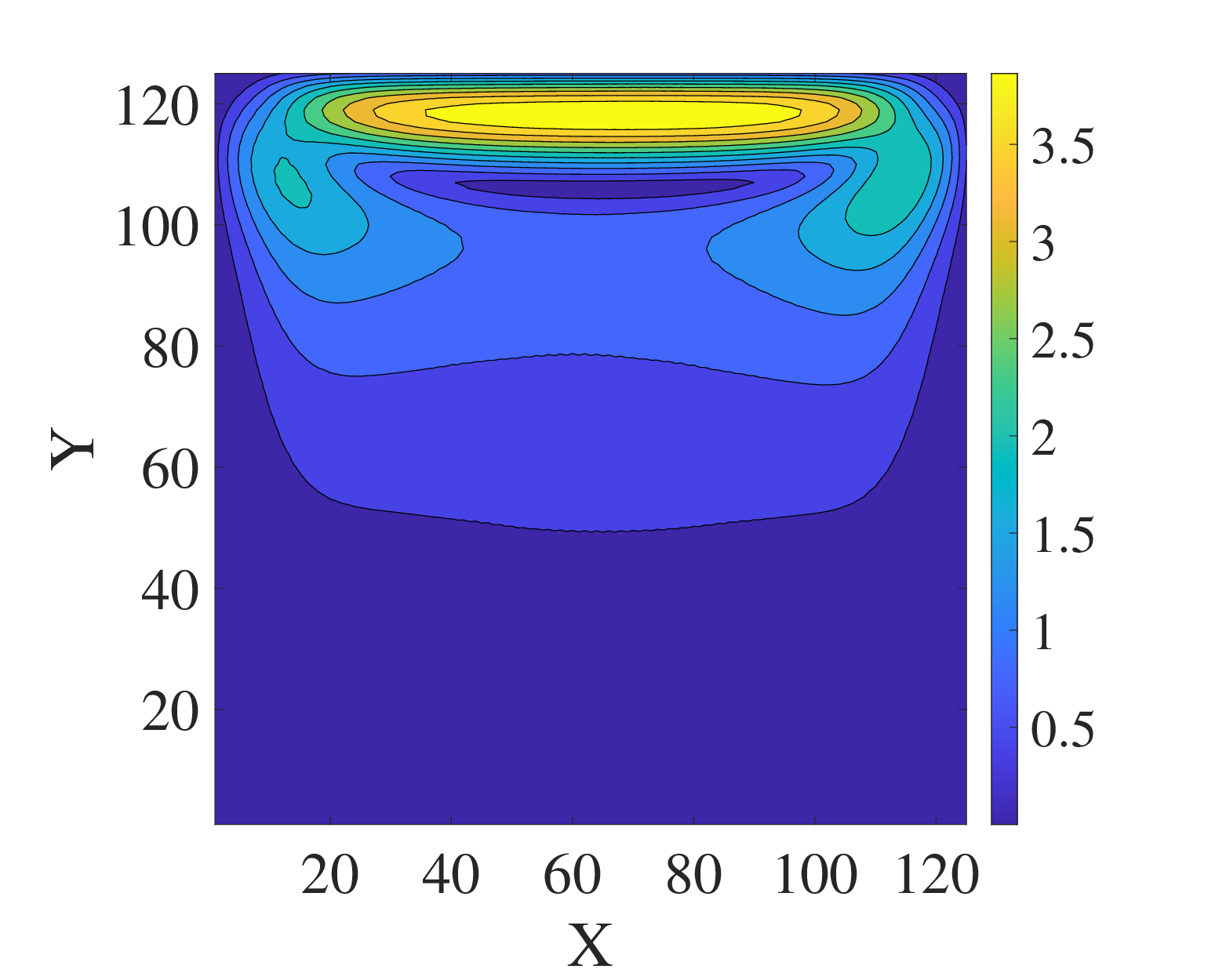}}
& \subfigure[Normalized Cost $\hat{\mathcal{A}}$.]{\label{fig:App2DPert}\includegraphics[width=6.5cm]{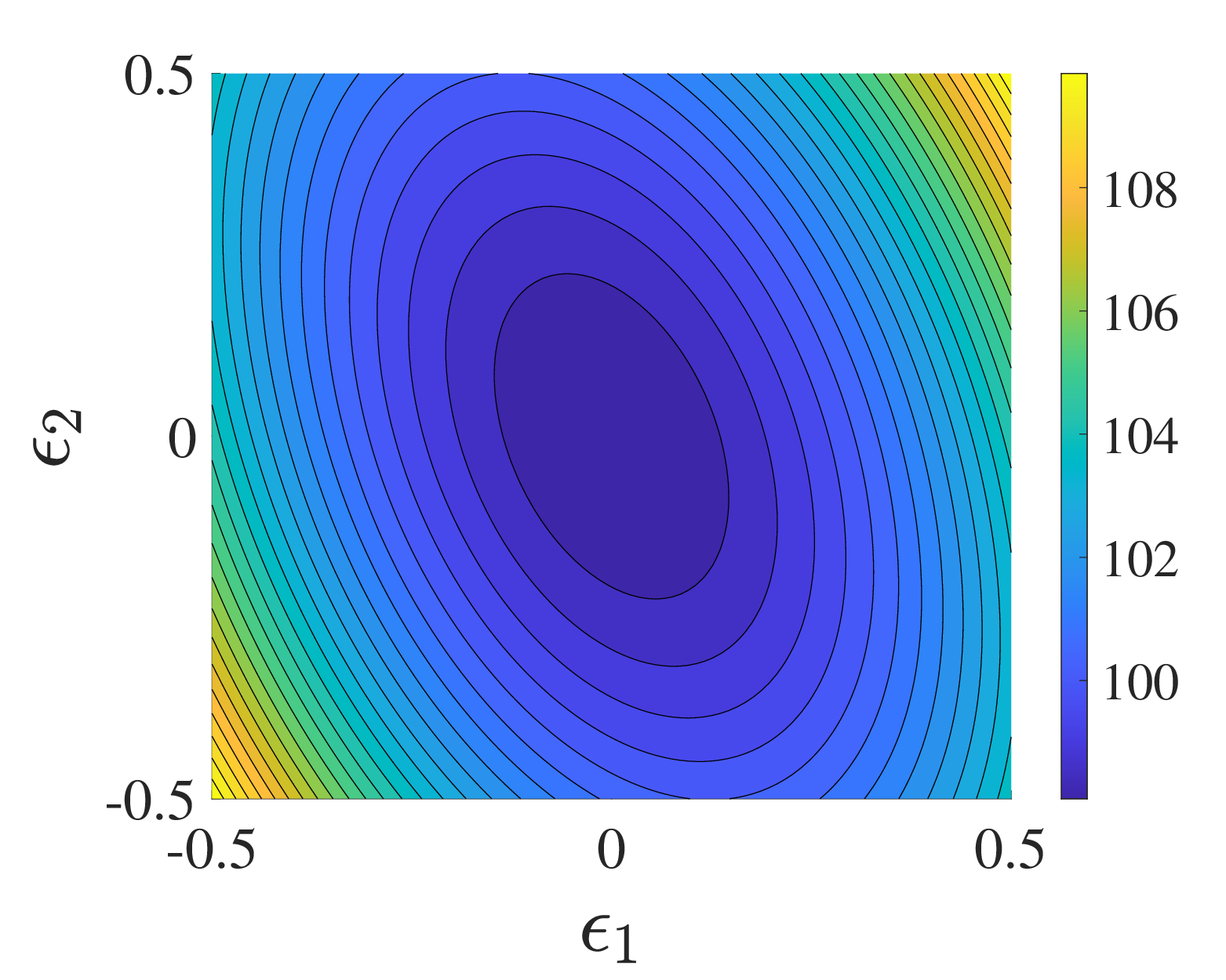}}

\end{array}$
\caption{Contours of the magnitude of the nondimensional evolution $\frac{\partial \bm{u}}{\partial t}\frac{T_{ref}}{U_{lid}}$ corresponding to (a, b) the perturbations $\bm\eta_1$, $\bm\eta_2$, and (c) the true evolution $\dot{\bm{U}}^*$. The subfigure (d) shows contours of the nondimensional cost $\hat{\mathcal{A}}$ in the $\epsilon_1$-$\epsilon_2$ plane. The pressure gradient cost attains its minimum precisely at $\epsilon_1=\epsilon_2=0$, confirming the optimality of $\dot{\bm{U}}^*$ over the family $\dot{\bm{U}} = \dot{\bm{U}}^* + \epsilon_1 \bm\eta_1 + \epsilon_2 \bm\eta_2$.}
\label{Fig:Minmiality_Demo2D}
\end{center}
\end{figure*}

Of course, one can obtain a similar behavior using a different perturbation direction $\bm\eta$ or even multiple perturbations. For instance, given two kinematically-admissible perturbations $\bm\eta_1$, $\bm\eta_2$, we construct the family of legitimate evolutions:
\[ \dot{\bm{U}} = \dot{\bm{U}}^* + \epsilon_1 \bm\eta_1 + \epsilon_2 \bm\eta_2\]
where $\epsilon_1$, $\epsilon_2$ control the perturbation magnitudes. We then compute the pressure gradient cost $\mathcal{A}(\dot{\bm{U}},\bm{U})$ corresponding to each member of the family. Figure \ref{Fig:Minmiality_Demo2D} shows contours of $\dot{\bm{U}}^*$, $\bm\eta_1$, and $\bm\eta_2$ in addition to the contours of the normalized cost $\hat{\mathcal{A}}$ in the plane $\epsilon_1$-$\epsilon_2$. The pressure gradient cost attains its minimum precisely at $\epsilon_1=\epsilon_2=0$, demonstrating the optimality of $\dot{\bm{U}}^*$.

\section{Conclusion}
In this paper, we adopt a pure analytical and variational mechanics approach to incompressible flows, fundamentally relying on Gauss's principle of least constraint. The principle asserts that a constrained mechanical system evolves at each instant in the closest possible manner to its \textit{free} motion---the motion that would occur in the absence of constraints. Since the deviation from the free motion is directly proportional to the \textit{constraint force} required to ensure the constraint, the principle is equivalent to minimizing the magnitude of the constraint force, hence the name least constraint. In other words, among all candidate motions that satisfy the given kinematical constraints, the true motion is the one that requires the smallest possible constraint force. This concept transforms the dynamics problem into a pure minimization problem, where the cost function is the total magnitude of the constraint force. Mathematically, Newton's equation of motion is the first-order necessary condition for minimizing this cost function. A recent extension of Gauss's principle to the continuum mechanics of incompressible flows led to the development of the Principle of Minimum Pressure Gradient (PMPG)  \citep{PMPG_PoF}. Analogous to Gauss's principle, the PMPG asserts that an incompressible flow evolves from one instant to another by minimizing the magnitude of the pressure gradient force required to ensure the continuity constraint. Equivalently, Navier-Stokes' equation is the first-order necessary condition for minimizing this pressure gradient cost function. Thus, the PMPG transforms incompressible fluid mechanics into a pure minimization problem, where the flow evolution is determined by minimizing a cost function---without explicitly invoking the Navier-Stokes equation.

Here, we demonstrate that the minimization problem induced by the PMPG is a convex Quadratic Programming (QP) problem: a nonlinear optimization problem with linear constraint and a quadratic cost whose Hessian is positive-definite. Fortunately, a convex QP problem is one of the most computationally tractable classes of nonlinear optimization. It has a unique solution that can be computed in polynomial time, supported by a rich literature and a well-established theoretical foundation with abundant efficient algorithms. The PMPG formulation eliminates the need to solve the Poisson equation in pressure at each time step; it replaces it with a convex QP problem. It remains, however, to perform a quantitative comparison between standard projection techniques that require solving a Poisson equation at each time step (or sub-step), and the presented convex QP formulation.

One of the key advantages of convex QP problems is the ability to obtain closed-form solutions in some cases. Exploiting tools from analytical mechanics and the Moore-Penrose theory of generalized inverses, we derive an explicit analytical solution of the QP problem. As such, we obtain an explicit formula for the projected dynamics of the spatially discretized Navier-Stokes equations on the space of divergence-free fields, eliminating the need for solving a Poisson equation in pressure or even numerically solving a QP problem. The result is an explicit ordinary differential equation (ODE) governing the evolution of the nodal values of the velocity field, which can be directly marched forward in time. We refer to this formulation as Variational Projection of Navier-Stokes (VPNS). Starting with a divergence-free velocity field, the VPNS guarantees that the resulting velocity field remains divergence-free at all future times. The VPNS formulation yields an explicit nonlinear ODE with constant coefficients in the case of fixed boundary conditions, which is expected to be quite valuable for simulation. Moreover, the compact form of the VPNS ODE will facilitate hydrodynamic stability analysis, flow control design, and the broader application of nonlinear systems theory, leveraging its rich legacy of analytical tools. We presented a proof-of-concept for the VPNS formulation by applying it to the benchmark problem of unsteady flow in a lid-driven cavity, and validating the resulting flow field against simulations in OpenFOAM.

\section*{Acknowledgments}
The first author is grateful to Dr. Ahmed Seleit for introducing the approach of Udwadia and Kalaba several years ago in a different context, related to optimal control.

\appendix

\section{Navier-Stokes Equation is the Necessary Condition for Optimality}\label{app:theorem}
\noindent\textbf{Theorem.} Let $\rho$ be a positive constant. Consider a candidate smooth ($C^\infty$) flow field $\bm{u}(\bm{x};t)$, defined over the domain $\Omega\subset\mathbb{R}^3$ with a smooth boundary $\partial\Omega$ and a time interval $t\in[0,T]$ for some $T>0$ such that: (i) $\bm{u}\cdot\bm\nabla \bm{u}$, $\bm\nabla \cdot \bm\tau$, $\bm{F}$ are in $\mathbb{L}^2(\Omega)$ for all $t\in[0,T]$, and (ii) the initial condition $\bm{u}(\bm{x};0)$ is kinematically-admissible:
\[ \bm\nabla\cdot\bm{u}(\bm{x};0)=0 \; \forall \; \bm{x}\in\Omega \;\; \mbox{and}\;\; \bm{u}(\bm{x};0)\cdot\bm{n}=g(\bm{x}) \; \forall \; \bm{x}\in\partial\Omega, \]
where $\bm{n}$ is the unit normal to the boundary $\partial\Omega$ and $g$ is a given smooth function on $\partial\Omega$.\\

If for every $t\in[0,T]$, the local acceleration $\bm{u}_t(\bm{x};t)\equiv \frac{\partial \bm{u}}{\partial t}(\bm{x};t)$ minimizes the functional
\[
 \mathcal{A}(\bm{u}_t) = \frac{1}{2}\int_\Omega \rho \left[\bm{u}_t+\bm{u}\cdot\bm\nabla \bm{u}-\frac{1}{\rho}\left(\bm\nabla\cdot\bm{\tau}+\bm{F}\right)\right]^2 d\bm{x} \]
over the space of admissible \textit{evolutions}:
\[ \Theta=\{\bm{u}_t \in\mathbb{L}^2| \; \bm\nabla \cdot\bm{u}_t=0 \; \rm{in} \; \Omega, \;\; \bm{u}_t\cdot \bm{n} = 0\; \rm{on} \; \partial\Omega\}, \]
then the candidate flow field $\bm{u}(\bm{x};t)$ must satisfy the Navier-Stokes equations
\[ \begin{array}{c}
\rho\left(\bm{u}_t+\bm{u}\cdot\bm\nabla \bm{u}\right)=-\bm\nabla p+\bm\nabla \cdot \bm\tau+\bm{F}, \\
\bm\nabla\cdot\bm{u}= 0 \end{array}\]
for all $\bm{x}\in\Omega$, $t\in[0,T]$, for some differentiable function $p$ on $\Omega\times[0,T]$, along with the no-penetration boundary condition
\[ \bm{u}(\bm{x};t)\cdot\bm{n}=g(\bm{x}) \; \forall \; \bm{x}\in\partial\Omega \;\; t\in[0,T]. \]

\[ \mathcal{L}(\bm{u}_t) = \mathcal{A}(\bm{u}_t) - \int_\Omega \lambda(\bm{x}) \left(\bm\nabla \cdot\bm{u}_t(\bm{x};t)\right) d\bm{x}, \]
where $\lambda$ is a Lagrange multiplier. A necessary condition for the constrained  minimization problem is then: the first variation of the Lagrangian vanishes with respect to variations in $\bm{u}_t(\bm{x})$ that belongs to $\Theta$. The first variation of $\mathcal{L}$ with respect to $\bm{u}_t$ is written as
\[\delta\mathcal{L} = \int_\Omega \left[\rho \left(\bm{u}_t+\bm{u}\cdot\bm\nabla \bm{u}-\bm\nabla \cdot \bm\tau-\bm{F} \right)\cdot\delta \bm{u}_t - \lambda \bm\nabla \cdot\delta\bm{u}_t \right] d\bm{x}=0.\]
The last term $\lambda \bm\nabla \cdot\delta\bm{u}_t$ can be written as
\[ \lambda \bm\nabla \cdot\delta\bm{u}_t = \bm\nabla \cdot (\lambda\delta\bm{u}_t)-\bm\nabla \lambda \cdot \delta\bm{u}_t \]
Integrating the first component $\bm\nabla \cdot (\lambda\delta\bm{u}_t)$ and using the divergence theorem, we have
\[ \int_\Omega \bm\nabla \cdot (\lambda\delta\bm{u}_t) d\bm{x} =  \int_{\partial\Omega} \lambda\delta\bm{u}_t\cdot\bm{n} d\bm{x}, \]

Since the variation $\delta\bm{u}_t$ belongs to $\Theta$, it must satisfy $\delta\bm{u}_t\cdot \bm{n}=0$ on the boundary, which results in $\int_\Omega \bm\nabla \cdot (\lambda\delta\bm{u}_t) d\bm{x}=0$. Hence, we have
\[ \delta\mathcal{L} = \int_\Omega \left[\rho \left(\bm{u}_t+\bm{u}\cdot\bm\nabla \bm{u}\right) + \bm\nabla \lambda - \bm\nabla \cdot \bm\tau-\bm{F} \right]\cdot \delta \bm{u}_t d\bm{x}=0 \]
for all admissible variations $\delta \bm{u}_t\in\Theta$, which implies
\[\rho\left(\bm{u}_t+\bm{u}\cdot\bm\nabla \bm{u}\right)=-\bm\nabla \lambda+\bm\nabla \cdot \bm\tau+\bm{F} \;\; \forall \; \bm{x}\in\Omega. \]
Moreover, this proof is valid for any $t\in[0,T]$.

In addition, since the initial condition $\bm{u}(\bm{x};0)$ is divergence-free for all $\bm{x}\in\Omega$ and the local acceleration $\bm{u}_t(\bm{x};t)$ is divergence-free for all $\bm{x}\in\Omega$ and $t\in[0,T]$, then it implies that
\[\bm\nabla\cdot \bm{u}(\bm{x},t) = 0 \;\; \forall\; \bm{x}\in\Omega \;\; \mbox{and} \; t\in[0,T].\]
Finally, since the initial condition $\bm{u}(\bm{x};0)$ satisfies the no-penetration boundary condition
\[ \bm{u}(\bm{x};0)\cdot\bm{n}=g(\bm{x}) \;\; \forall \bm{x}\in\partial\Omega,\]
and the local acceleration $\bm{u}_t(\bm{x};t)$ satisfies a homogenous normal boundary condition
\[ \bm{u}_t(\bm{x};t)\cdot\bm{n}=0 \;\; \forall \bm{x}\in\partial\Omega \;\; \mbox{and} \; t\in[0,T],\]
then it implies that
\[ \bm{u}(\bm{x};t)\cdot\bm{n}=g(\bm{x}) \;\; \forall \bm{x}\in\partial\Omega \;\; \mbox{and} \; t\in[0,T],\]
which concludes our proof. $\blacksquare$

\section{Setup of the QP Problem for the Unsteady Lid-Driven Cavity}\label{appA}
\subsection{Spatial Discretization}
\begin{figure*}
\begin{center}
$\begin{array}{cc}
\subfigure[The Cavity Square Domain.]{\label{fig:ldcDomain}\includegraphics[width=6.5cm]{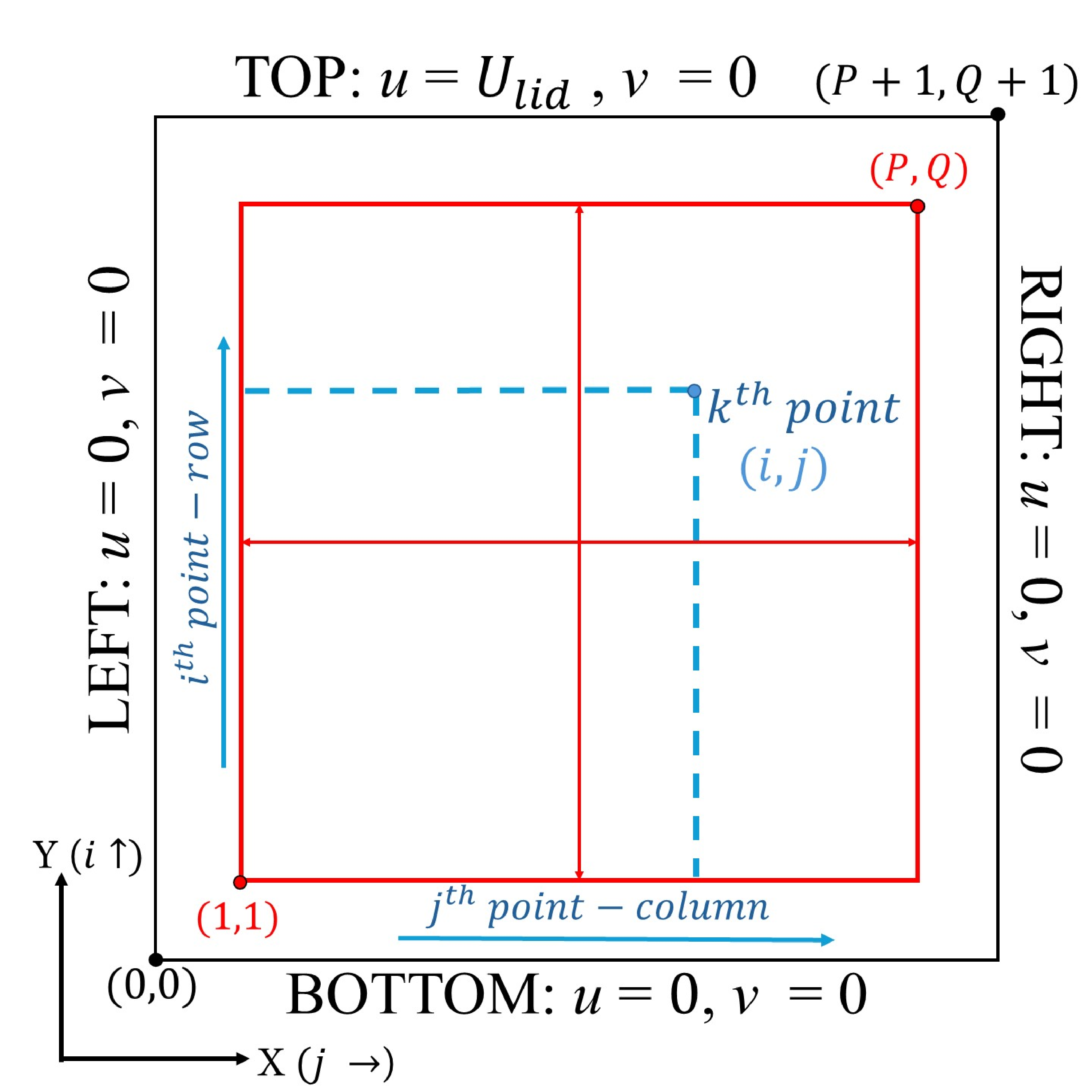}} & \subfigure[Five-Point Stencil.]{\label{fig:stencil2D}\includegraphics[width=6.5cm]{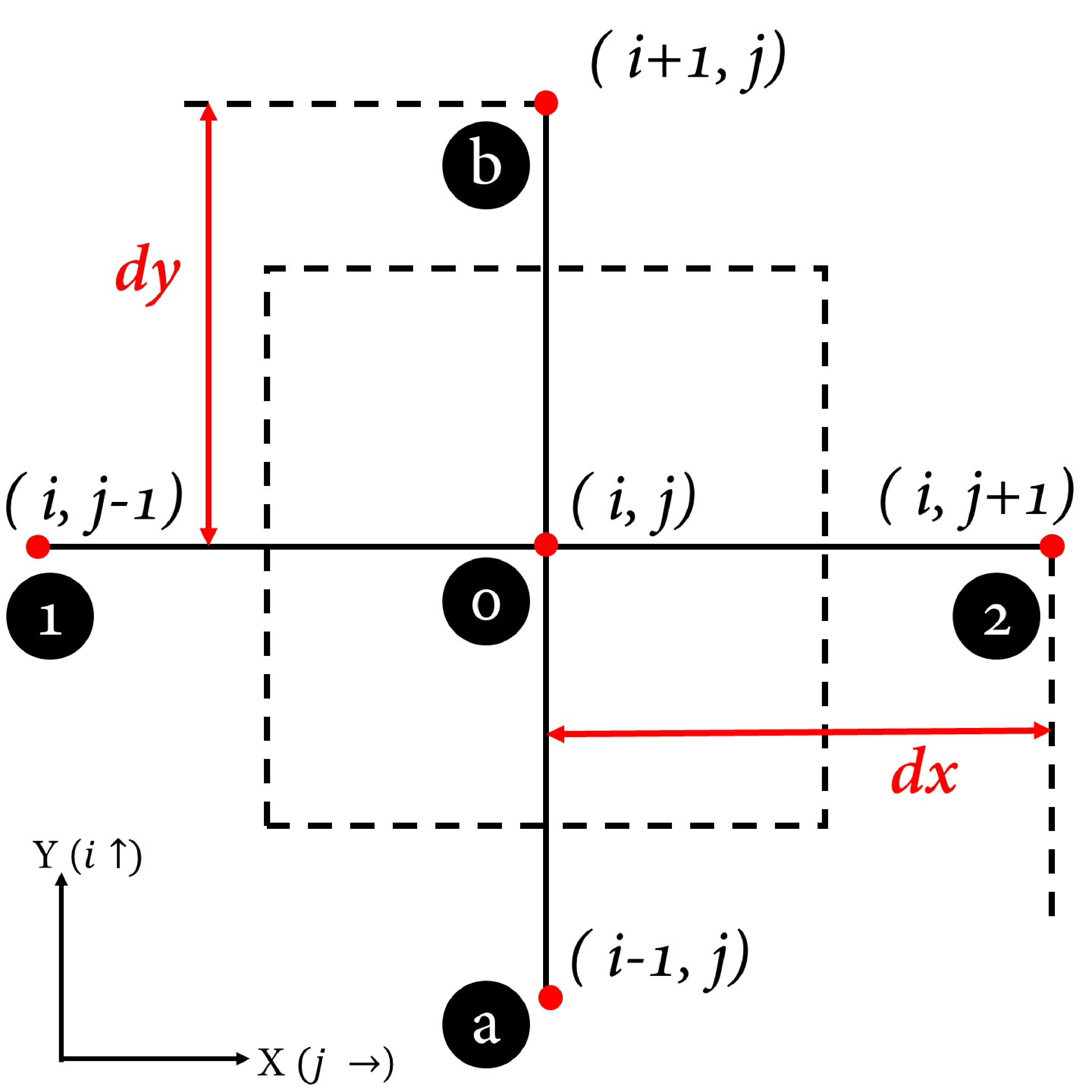}}
\end{array}$
\caption{A schematic for the square domain of the lid-driven cavity problem along with the five-point stencil for our grid.}
\label{Fig:Schematic}
\end{center}
\end{figure*}

The square domain is divided into a regular grid of $P\times Q$ interior points, as shown in Fig. \ref{fig:ldcDomain}. We use a five-point stencil (Fig. \ref{fig:stencil2D}) for spatial discretization with three-point central differencing. For simplicity, we use the same discretization for both the convective and viscous terms. Although it is generally not preferable to use three point central differencing scheme with the convective term, it is a reasonable approximation for low Peclet numbers. It should be noted that QP formulation (\ref{eq:PMPG_QP_Problem}) can be constructed using any standard discretization method and is not limited to the specific discretization choices made in this demonstration.

The cost functional $\mathcal{A}$ is written as a summation over the elements
\[ \mathcal{A} = \sum_{i,j=1}^{P,Q} \mathcal{A}_{i,j}, \]
where
\begin{equation}\label{eq:Cost_Element_Basic}
    \mathcal{A}_{i,j} = \frac{1}{2}m_{i,j}\left( \bm{u}_{t_{i,j}} + \bm{u}_{i,j} \cdot \bm\nabla \bm{u}_{i,j} - \nu \bm\nabla^2 \bm{u}_{i,j}\right)^2,
\end{equation}
where $m_{i,j} = \rho \, \Omega_{i,j}$, $\Omega_{i,j}$ is the area of the element $(i,j)$, and a Newtonian viscous stress model is used with a kinematic viscosity coefficient $\nu$. Let
\[ \bm{u}_{i,j} = u_0 \hat{x} + v_0 \hat{y} \;\;\mbox{and}\;\; \bm{u}_{t_{i,j}} = \dot{u_0} \hat{x} + \dot{v_0} \hat{y}, \]
where the labels $0,1,2$, and $a,b$ follow Fig. \ref{fig:stencil2D}, and $\hat{x}$, $\hat{y}$ are unit vectors along the $x$ and $y$ axes, respectively. The convective term is then given by
\[
\begin{aligned}
\left(\bm{u} \cdot \bm\nabla \bm{u}\right)_{i,j} 
&= \underbrace{\left( \frac{u_0 (u_2 - u_1)}{2 dx} + \frac{v_0 (u_b - u_a)}{2 dy} \right)}_{\substack{\text{Term A}}} \hat{x} \\
&\quad + \underbrace{\left( \frac{u_0 (v_2 - v_1)}{2 dx} + \frac{v_0 (v_b - v_a)}{2 dy} \right)}_{\substack{\text{Term B}}} \hat{y},
\end{aligned}
\]

And the discretized viscous term is written as:
\[
\begin{aligned}
\nu \bm\nabla^2 \bm{u}_{i,j} 
&= \underbrace{\nu \left( \frac{u_2 - 2 u_0 + u_1}{dx^2} + \frac{u_b - 2 u_0 + u_a}{dy^2} \right)}_{\substack{\text{Term } \alpha}} \hat{x} \\
&\quad + \underbrace{\nu \left( \frac{v_2 - 2 v_0 + v_1}{dx^2} + \frac{v_b - 2 v_0 + v_a}{dy^2} \right)}_{\substack{\text{Term } \beta}} \hat{y}
\end{aligned}
\]

As such, the element cost $\mathcal{A}_{i,j}$ is written in terms of the terms $A$, $B$, $\alpha$ and $\beta$, defined above, as:
\[
\mathcal{A}_{i,j}  = \frac{1}{2} m_{i,j}\left( \dot{u}_0 + A - \alpha \right)^2 + \frac{1}{2} m_{i,j}\left( \dot{v}_0 + B - \beta \right)^2,
\]
which yields
\[
\begin{aligned}
\mathcal{A}_{i,j} 
&= \underbrace{\frac{1}{2} m_{i,j}\left( \dot{u}_0^2 + \dot{v}_0^2 \right)}_{\substack{\text{Term I}}} \\
&\quad + \underbrace{m_{i,j} \bigg[\dot{u}_0 (A - \alpha) + \dot{v}_0 (B - \beta)\bigg]}_{\substack{\text{Term II}}} \\
&\quad + \underbrace{\frac{1}{2} m_{i,j} \bigg[A^2 + \alpha^2 - 2 A \alpha + B^2 + \beta^2 - 2 B \beta \bigg]}_{\substack{\text{Term III}}}
\end{aligned}
\]

Each of these terms is written in a matrix-form as:



\[
\begin{array}{lll}
\text{Term I} &=& \frac{1}{2}
\begin{bmatrix}
\dot{u}_0 & \dot{v}_0
\end{bmatrix}
\begin{bmatrix}
m_{i,j} & 0 \\
0 & m_{i,j}
\end{bmatrix}
\begin{bmatrix}
\dot{u}_0 \\
\dot{v}_0
\end{bmatrix} \\[1ex]

\text{Term II} &=& 
\begin{bmatrix}
m_{i,j} (A - \alpha) & m_{i,j} (B - \beta)
\end{bmatrix}
\begin{bmatrix}
\dot{u}_0 \\
\dot{v}_0
\end{bmatrix} \\[1ex]

\text{Term III} &=& \frac{1}{2}
\begin{bmatrix}
m_{i,j} (A - \alpha) & m_{i,j} (B - \beta)
\end{bmatrix} \\[0.5ex]
&& \quad
\begin{bmatrix}
\frac{1}{m_{i,j}} & 0 \\
0 & \frac{1}{m_{i,j}}
\end{bmatrix}
\begin{bmatrix}
m_{i,j} (A - \alpha) \\
m_{i,j} (B - \beta)
\end{bmatrix}
\end{array}
\]

Thus, letting
\begin{equation}\label{eq:Mass_Matrix_Element}
    \dot{\bm{U}}_{i,j} =
\begin{pmatrix}
\dot{u}_0 \\
\dot{v}_0
\end{pmatrix}
, \quad
\bm{M}_{i,j} =
\begin{pmatrix}
m_{i,j} & 0 \\
0 & m_{i,j}
\end{pmatrix}
, \quad
\dot{\bm{U}}^{\rm{free}}_{i,j} =
\begin{pmatrix}
 \alpha-A \\
 \beta-B
\end{pmatrix},
\end{equation}
the element cost $\mathcal{A}_{i,j}$ can be written as
\begin{equation}\label{eq:Cost_Element_Matrix}
\mathcal{A}_{i,j}=\frac{1}{2} \left( \dot{\bm{U}}_{i,j} - \dot{\bm{U}}^{\rm{free}}_{i,j} \right)^T
\bm{M} \left( \dot{\bm{U}}_{i,j} - \dot{\bm{U}}^{\rm{free}}_{i,j} \right),
\end{equation}

The continuity constraint $\bm\nabla \cdot\dot{\bm{u}}=0$, imposed on the element $(i,j)$, is given by
\begin{equation}\label{eq:Constraint_Element_Matrix}
\frac{\dot{u}_2 - \dot{u}_1}{2dx} + \frac{\dot{v}_b - \dot{v}_a}{2dy} = 0 \;\; \Rightarrow \;\; \bigg[\frac{-1}{2dy} \,\, \frac{-1}{2dx} \,\, \frac{1}{2dx} \,\, \frac{1}{2dy}\bigg] \begin{pmatrix}
        \dot{v}_a \\
        \dot{u}_1 \\
        \dot{u}_2 \\
        \dot{v}_b
    \end{pmatrix} = 0.
\end{equation}

\subsection{Domain Assembly}
Having constructed the cost (\ref{eq:Cost_Element_Matrix}) and constraint (\ref{eq:Constraint_Element_Matrix}) for each element, the next step is to perform assembly for the whole domain. Since both components of the velocity vector are specified on all boundaries, the array $\bm{U}$ of unknown nodal values includes only interior points. The components of the velocity vector at each one of these points are stacked in the array as:
\begin{equation}\label{eq:Stacking}
    \bm{U} = [u_{1,1} \,\, v_{1,1} \,\, u_{1,2} \,\, v_{1,2} \,\, \ldots \,\, u_{P,Q} \,\, v_{P,Q}]^T \in\mathbb{R}^{2PQ}
\end{equation}
The velocity vector is arranged in a row wise flattened manner over the grid starting from the base of the grid. Consequently, the interior point $(i,j)$ will have the global location $k$ in the array $\bm{U}$, where
\[
    k = j + (i-1)Q \;\; \mbox{and} \;\; u_k = \bm{U}(2k-1), \;\; v_k = \bm{U}(2k).
\]

The mass matrix $\bm{M}$ is easily constructed by stacking the individual mass matrices $\bm{M}_{i,j}$, defined in Eq. (\ref{eq:Mass_Matrix_Element}), on the diagonal:
\[ \bm{M} = \left[\rm{diag}(\bm{M}_k)\right]\in\mathbb{R}^{2PQ\times 2PQ}.\]
However, it is worth mentioning that, for practical implementation, we need \(\bm{M}^{-\frac{1}{2}}\) and since \(\bm{M }\) is a diagonal matrix, we rather compute the inverse of the square root of the diagonal elements of \(\bm{M }\) and form a diagonal matrix with these numbers instead.

To construct the global array $\dot{\bm{U}}^{\rm{free}}\in\mathbb{R}^{2PQ}$ of free accelerations from the local one $\dot{\bm{U}}^{\rm{free}}_k\in\mathbb{R}^2$, given in Eq. (\ref{eq:Mass_Matrix_Element}), we recall the two components of $\dot{\bm{U}}^{\rm{free}}_k$:

\[
\begin{array}{lll}
\alpha - A &=& 
-\left[ \frac{u_0 (u_2 - u_1)}{2dx} + \frac{v_0 (u_b - u_a)}{2dy} \right] \\[0.5ex]
&& \quad + \nu \left[ \frac{u_2 - 2u_0 + u_1}{dx^2} + \frac{u_b - 2u_0 + u_a}{dy^2} \right], \\[1.5ex]

\beta - B &=& 
-\left[ \frac{u_0 (v_2 - v_1)}{2dx} + \frac{v_0 (v_b - v_a)}{2dy} \right] \\[0.5ex]
&& \quad + \nu \left[ \frac{v_2 - 2v_0 + v_1}{dx^2} + \frac{v_b - 2v_0 + v_a}{dy^2} \right],
\end{array}
\]

which require the definitions

\[
\begin{array}{lllll}
u_0 = \bm{U}(2k-1), & u_1 = \bm{U}(2k-3), \\[0.4ex]
u_2 = \bm{U}(2k+1), & u_a = \bm{U}(2k-2Q-1), \\[0.4ex]
u_b = \bm{U}(2k+2Q-1), & & & & \\[1.2ex]

v_0 = \bm{U}(2k), & v_1 = \bm{U}(2k-2), \\[0.4ex]
v_2 = \bm{U}(2k+2), & v_a = \bm{U}(2k-2Q), \\[0.4ex]
v_b = \bm{U}(2k+2Q), & & & &
\end{array}
\]

for all $k = j + (i-1)Q$, $i\in\{2,...P-1\}$ and $j\in\{2,...Q-1\}$; i.e., excluding the points adjacent to the boundaries shown in red in Fig. \ref{fig:ldcDomain}. We will show below how to handle these points as part of the boundary conditions because their free accelerations are directly impacted by them. Once each $\dot{\bm{U}}^{\rm{free}}_k$ is constructed, they are stacked in a similar fashion to $\bm{U}_k$ in Eq. (\ref{eq:Stacking}) in order to obtain $\dot{\bm{U}}^{\rm{free}}\in\mathbb{R}^{2PQ}$.

Similarly, the elements of the row, defined in Eq. (\ref{eq:Constraint_Element_Matrix}) for the continuity constraint on the interior point $(i,j)$, can be used to construct the $k^{\rm{th}}$ row of the global constraint matrix $\bm{D}\in\mathbb{R}^{PQ}$ as
\begin{equation}\label{eq:Global_Constraint_Matrix}
\bm{D}(k,\ell)=\left\{\begin{array}{cl}
-\frac{1}{2dy}, & \ell=2k-2Q \\
-\frac{1}{2dx}, & \ell=2k-3\\
\frac{1}{2dx}, & \ell=2k+1\\
\frac{1}{2dy}, & \ell=2k+2Q\\
0 & \mbox{Otherwise}. \end{array}\right.
\end{equation}
for all $i\in\{2,...P-1\}$ and $j\in\{2,...Q-1\}$.

For the relatively simple example of laminar flow in a lid-driven cavity, considered in this work, the Moore-Penrose inverse of the discrete divergence operator \( \tilde{\mathbf{D}} \) is computed using the identity \( \tilde{\mathbf{D}}^+ = \tilde{\mathbf{D}}^\top (\tilde{\mathbf{D}} \tilde{\mathbf{D}}^\top)^{-1} \), relying on the full row rank structure of \( \tilde{\mathbf{D}} \). Numerical symmetry and mass-conserving properties of the resulting projection operator \( \mathbf{N} = \mathbf{I} - \tilde{\mathbf{D}}^+ \tilde{\mathbf{D}} \) are preserved by careful discretization.

\subsection{Boundary Conditions}
The given Dirichlet boundary conditions on $u$ and $v$ are enforced by excluding the boundary points from the array $\bm{U}$ of unknown nodal values; i.e., the points $(i,j)$ with $i=0$, $i=P+1$, $j=0$, or $j=Q+1$. Additionally, the given values of $u$, $v$ at these points affect the free accelerations and the continuity constraint at the neighboring points---the points $(i,j)$ with $i=1$, $i=P$, $j=1$, or $j=Q$, which are shown in red in Fig. \ref{fig:ldcDomain}. For these points, we have
\[ \begin{array}{lll} u_1=0, & v_1=0, &  j=1 \\
u_2=0, & v_2=0, & j=Q \\
u_a=0, & v_a=0, & i=1 \\
u_b=U_{lid}, & v_b=0, & i=P.\end{array} \]
Consequently, the array $\dot{\bm{U}}^{\rm{free}}$ of free accelerations will depend on the nodal values $\bm{U}$ as well as a term $\bm{B}$ which is dictated by the boundary conditions:
\[ \dot{\bm{U}}^{\rm{free}} = \bm{C}(\bm{U}) + \bm{B}, \]
where $\bm{C}(.)$ is a quadratic form in $\bm{U}$ because of the convective term, and $\bm{B}$ is a constant array for fixed boundary conditions.

Also, the $k^{\rm{th}}$ row of the constraint matrix $\bm{D}$ for $k = j + (i-1)Q$ with $i=1$, $i=P$, $j=1$, or $j=Q$ is affected by the boundary conditions. If $u$ and $v$ at the boundary points are not time-varying, then we have
\[ \begin{array}{lll} \dot{u}_1=0, & \dot{v}_1=0, &  j=1 \\
\dot{u}_2=0, & \dot{v}_2=0, & j=Q \\
\dot{u}_a=0, & \dot{v}_a=0, & i=1 \\
\dot{u}_b=0, & \dot{v}_b=0, & i=P,\end{array} \]
which implies that these rows may follow the definition (\ref{eq:Global_Constraint_Matrix}) with overriding some of the elements as follows: $\bm{D}(k,\ell)=0$ if

\[
\begin{array}{lll}
k = 1 + (i-1)Q, \quad \ell = 2(i-1)Q - 1, \\
k = iQ, \quad \ell = 2iQ - 1, & \forall \; i \in \{1, \ldots, P\}, \\[1.2ex]

k = j, \quad \ell = 2j - 2Q, \\
k = j + (P-1)Q, \quad \ell = 2j + 2PQ, & \forall \; j \in \{1, \ldots, Q\}
\end{array}
\]

We have now constructed the main ingredients of the QP problem (\ref{eq:PMPG_QP_Problem}): $\bm{M}$, $\bm{D}$, and $\dot{\bm{U}}^{\rm{free}}$.

\bibliographystyle{apsrev4-2}
\bibliography{Merged_References}

\end{document}